\documentclass[twocolumn,showpacs,preprintnumbers,amsmath,amssymb,showkeys]{revtex4}

\usepackage{graphicx}
\usepackage{dcolumn}
\usepackage{bm}
\usepackage{bezier}
\usepackage{color}

\newcommand{\nag}{\phantom{\dag}}

\begin{document}

\title{Phonon-affected steady-state transport through molecular quantum dots}
\author{T. Koch, H. Fehske}
\affiliation{Institut f{\"u}r Physik,
            Ernst-Moritz-Arndt-Universit{\"a}t Greifswald,
             DE-17489 Greifswald, Germany
}
\email{thomas.koch@physik.uni-greifswald.de}
\author{J. Loos}
\affiliation{Institute of Physics, Academy of Sciences of the Czech Republic, CZ-16200 Prague, Czech Republic
}

\date{\today}

\begin{abstract}
We consider transport through a vibrating molecular quantum dot contacted to macroscopic leads acting as charge reservoirs. In the equilibrium and nonequilibrium regime, we study the formation of a polaron-like transient state at the quantum dot for all ratios of the dot-lead coupling to the energy of the local phonon mode. We show that the polaronic renormalization of the dot-lead coupling is a possible mechanism for negative differential conductance. Moreover, the effective dot level follows one of the lead chemical potentials to enhance resonant transport, causing novel features in the inelastic tunneling signal. In the linear response regime, we investigate the impact of the electron-phonon interaction on the thermoelectrical properties of the quantum dot device.
\end{abstract}
\maketitle
%
%
\section{Introduction}

\label{sec:introduction}
Electronic devices featuring a single organic molecule as the active element, so called molecular junctions, are promising candidates in the search for further miniaturization and novel functionality. Such systems can be described as quantum dots: mesoscopic systems coupled to macroscopic charge and heat reservoirs. 

Molecular junctions are susceptible to structural changes when being occupied by charge carriers. 
The local interaction with optical phonons becomes apparent as vibrational signatures in the current-voltage characteristics of the device~\cite{HM93,Reea97,Paea00}, resulting from the interference of elastic and inelastic tunneling processes and the renormalization of the effective dot level energy~\cite{Pe88,MTU02,MTU03,GRN04,GRN07}. When the vibrational energy and the electron-phonon (EP) interaction become sufficiently large, nonlinear phenomena emerge, such as hysteresis, switching and negative differential conductance (NDC). As is well known from the Holstein molecular crystal model~\cite{Ho59a,Ho59b}, strong EP interaction may heavily reduce the ``mobility" of electrons through the formation of small polarons~\cite{WF97,WF98a,FT07,FWLB08}. Thus, the formation of a local polaron is considered a possible mechanism for the observed nonlinear transport properties of molecular junctions~\cite{LD07}.

Molecular junctions may also constitute efficient power generators or heat pumps. Their highly energy-dependent transmission together with the tunable level energy could be used to optimize the thermoelectrical figure of merit. In the weak dot-lead (DL) coupling limit, the theoretical efficiency approaches the Carnot value~\cite{HNTL02}. However, long electron residence times increase the effective EP coupling. Moreover, some level broadening is needed to ensure useable power output. That is why, for practical applications, the regime of comparable electronic and phononic time scales becomes interesting.

In our work, we calculate the steady-state charge and energy transport through the quantum dot for small-to-large DL coupling and weak-to-strong EP interaction. Based on a variational Lang-Firsov transformation~\cite{LF62,Feea94,LD07,LKABF09,KLABF10,KLAF11}, we determine the nonequilibrium dot spectral function in the formalism of Kadanoff-Baym~\cite{KB62} and calculate the dot self-energy in a self-consistent way up to second order in the renormalized interaction coefficients. The variational parameter is determined numerically by minimizing the thermodynamic potential.
%
%
\section{Model}
\label{sec:model}
We consider the standard Hamiltonian of the single-site quantum dot. It is based on a modified Fano-Anderson model with the static impurity being replaced by a single site coupled to a local phonon mode ($\hbar=1$):
\begin{align}
 H  &=  (\Delta-\mu) d^\dag d^{\nag} -\; g \omega_0 d^\dag d ( b^{\dag} + b) + \omega_0  b^{\dag} b\label{EQUhamiltonianstart} \\
& + \sum_{k,a} (\varepsilon_{ka}^{\nag}-\mu) c_{ka}^{\dag} c_{ka}^{\nag} - \frac{1}{\sqrt{N}}\sum_{k, a}\left(t_{ka}d^{\dag}c_{ka}^{\nag} + t_{ka}^\ast c_{ka}^{\dag}d\right) .\nonumber
\end{align}
The quantum dot is represented by the energy level $\Delta$, with the fermionic operators $d^{(\dag)}$. It is coupled to a local phonon mode $b^{(\dag)}$ of energy $\omega_0$, with the dimensionless EP coupling strength $g$. The operators $c_{ka}^{(\dag)}$ (for $k=1,\dots,N$; $a=L,R$) correspond to free electrons in the $N$ states of the left and right lead, with the energies $\varepsilon_{ka}$ and the equilibrium chemical potential $\mu$. The last term in Eq.~(\ref{EQUhamiltonianstart}) allows for dot-lead particle transfer.

To account for the competition between polaron localization and charge transport, we apply to the model (\ref{EQUhamiltonianstart}) an incomplete Lang-Firsov transformation~\cite{KLAF11}, introducing the variational parameter $\gamma\in[0,1]$. Then $\widetilde H=S_\gamma^\dag H S_\gamma$, with
\begin{align}
S_\gamma&=\exp\{g (b^{\dag}-b)(\gamma d^\dag d +(1-\gamma)n_d)\}\;.
\end{align}
For $\gamma=1$, $S_\gamma$ coincides with the shift-transformation of the Lang-Firsov small-polaron theory~\cite{LF62}, which eliminates the EP coupling term in Eq.~(\ref{EQUhamiltonianstart}) and  lowers the dot level by the polaron binding energy $\varepsilon_p=g^2\omega_0$. For $\gamma<1$, $S_\gamma$ accounts for the quasistatic displacement of the equilibrium position of the oscillator, which is proportional to the dot mean occupation $n_d=\langle d^{\dag}d \rangle$.
After the transformation the Hamiltonian reads
\begin{align}
\widetilde H &=  \widetilde\eta\, d^\dag d^{\nag} -g\omega_0(1-\gamma)(b^\dag+b) (d^\dag d-n_d) \nonumber \\
& + \omega_0  b^{\dag} b + \varepsilon_p(1-\gamma)^2n_d^2 +\sum_{k,a} (\varepsilon_{ka}^{\nag}-\mu)c_{ka}^{\dag} c_{ka}^{\nag}\label{EQUhamiltonian}\\
& -\frac{1}{\sqrt{N}}\sum_{k,a} \left(  t_{ka}\mathrm{e}^{-\gamma g (b^{\dag}-b)} d^{\dag}c_{ka}^{\nag}+ t_{ka}^\ast\mathrm{e}^{\gamma g (b^{\dag}-b)}c_{ka}^{\dag}d\right)  \;.\nonumber
\end{align}
In (\ref{EQUhamiltonian}), the DL coupling is affected by the EP interaction. Furthermore the bare dot level is renormalized:
\begin{align}
\widetilde\eta&=\Delta-\mu-\varepsilon_p\gamma(2-\gamma)-2\varepsilon_p(1-\gamma)^2n_d\;.\label{EQUdefeta}
\end{align}
Note that now $d$ and $b$ are the operators of dressed electrons (in analogy to polarons) and the shifted oscillator.
The original electron and oscillator operators now read $\widetilde d= \mathrm{exp}\{\gamma g (b^{\dag}-b)\}\,d$ and $\widetilde b=b+\gamma g d^\dag d+(1-\gamma) g n_d$.
 
The application of a potential difference between the leads is described by adding to (\ref{EQUhamiltonian}) the interaction with the external fields $U_a=-\delta\mu_a $ and defining the voltage bias $\Phi$, with $e$ being the negative elementary charge:
\begin{align}\label{EQUhint}
H_{\mathrm{int}}&=\sum_{a}U_a\sum_{k}c_{ka}^\dag c_{ka}^{\nag}\;,\quad\Phi&=(U_L-U_R)/e \;.
\end{align}
%
%
%
\section{Theoretical approach}
\label{sec:theoapp}
\subsection{Polaronic spectral function in the Kadanoff-Baym formalism}
For finite voltage bias between the noninteracting macroscopic leads, the response of the quantum dot is given by the polaronic nonequilibrium real-time Green functions
\begin{align}
g_{dd}^<(t_1,t_2;U)&=\mathrm{i}\langle  d_U^\dagger(t_2)d_U(t_1)\rangle \;,\label{EQUdefresponseless}\\
g_{dd}^>(t_1,t_2;U)&=-\mathrm{i}\langle  d_U(t_1)d_U^\dagger(t_2)\rangle \;,\label{EQUdefresponsegtr}
\end{align}
where the time dependence of $d_U^{(\dagger)}$ is determined by $\widetilde H+H_{\mathrm{int}}$. According to Kadanoff-Baym~\cite{KB62}, the real-time response functions may be deduced using the equations of motion for the nonequilibrium Green functions $G_{dd}^{\gtrless}(t_1,t_2;U,t_0)$ of the complex time variables $t=t_0-\mathrm{i}\tau$, $\tau\in[0,\beta]$. We base our calculations on the Dyson equation of the polaronic Green functions, which defines the polaronic self-energy $\Sigma_{dd}=G_{dd}^{(0)-1}-G_{dd}^{-1}\,$. For a given ordering of $t_1$, $t_2$, the equations of motion of the functions $g_{dd}^{\gtrless}$ follow through the limiting procedure $t_0\to -\infty$. Limiting ourselves to the steady-state regime, we suppose that all functions depend only on $t=t_1-t_2$. After a Fourier transformation by the method used in Ref.~\onlinecite{KB62}, the following exact equations for the steady-state are obtained~\cite{KLAF11}:
\begin{align}
g_{dd}^<(\omega;U)\Sigma_{dd}^>(\omega;U) -g_{dd}^>(\omega;U)&\Sigma_{dd}^<(\omega;U) = 0\;,\label{EQUsteady1}\\[0.2cm]
\left[\omega-\widetilde\eta-\mathrm{Re}\;\Sigma_{dd}(\omega;U)\right]A(\omega;U)& =\nonumber\\ \Gamma(\omega;U)&\;\mathrm{Re}\;g_{dd}(\omega;U)\;.\label{EQUsteady2}
\end{align}
Here we defined, in analogy to the equilibrium case,
\begin{align}
A(\omega;U) &= g_{dd}^>(\omega;U)+g_{dd}^<(\omega;U)\;,\label{EQUdefa} \\
g_{dd}(z;U)&=\int \frac{\mathrm{d}\omega}{2\pi}\; \frac{A(\omega;U)}{z-\omega}\;,\label{EQUdefg}\\
\Gamma(\omega;U)&=\Sigma_{dd}^>(\omega;U)+\Sigma_{dd}^<(\omega;U) \;,\label{EQUdefGamma}\\ 
\Sigma_{dd}(z;U)&=\int\frac{\mathrm{d}\omega}{2\pi}\;\frac{\Gamma(\omega;U)}{z-\omega}\;,\label{EQUdefSigma}
\end{align}
where $A(\omega;U)$ is the polaronic nonequilibrium spectral function. According to Eq.~(\ref{EQUdefa}), we can write
\begin{align}
g_{dd}^<(\omega;U)&=A(\omega;U)\bar f(\omega;U)\;,\label{EQUgless}\\
g_{dd}^>(\omega;U)&=A(\omega;U)(1- \bar f(\omega;U))\;,\label{EQUggtr}
\end{align}
introducing the nonequilibrium distribution $ \bar f$, which follows from Eqs. (\ref{EQUsteady1}) and (\ref{EQUdefGamma}) as
\begin{equation}\label{EQUbarf}
 \bar f (\omega;U)= \frac{\Sigma_{dd}^<(\omega;U)}{\Gamma(\omega;U)}\;.
\end{equation}
For the Green function $g_{dd}$ in Eq.~(\ref{EQUdefg}) we use the ansatz $g_{dd}(z;U)=1/(z-\widetilde\eta-\Sigma_{dd}(z;U))$ and find the following formal solution of Eq.~(\ref{EQUsteady2}):
\begin{align}\label{EQUafrac}
\hspace{-0.2cm}A(\omega;U)=\frac{\Gamma(\omega;U)}{\left(\omega-\widetilde\eta-\mathcal{P}\int\frac{\mathrm{d}\omega^\prime}{2\pi}\;\frac{\Gamma(\omega^\prime;U)}{\omega-\omega^\prime}\right)^2+\left(\frac{\Gamma(\omega;U)}{2}\right)^2}\;.
\end{align}
To deduce a functional differential equation for the self-energy $\Sigma_{dd}$ we add to $H_\mathrm{int}$ in Eq.(\ref{EQUhint}) the interaction with fictitious external fields $\{V\}$ (cf. Refs. \onlinecite{KB62,LKABF09,KLABF10,KLAF11,Sc66}). The equations of motion of the polaronic Green functions are then expressed by means of the functional derivatives of $\Sigma_{dd}$ with respect to $\{V\}$. The resulting equations for $\Sigma_{dd}^{\gtrless}$ are solved iteratively to the second order in the renormalized EP and DL interaction coefficients in (\ref{EQUhamiltonian}), while the correlation functions of the interaction coefficients are evaluated supposing independent Einstein oscillators. We then let $\{V\}\to 0$ and perform the limit $t_0\to -\infty$. A subsequent Fourier transformation yields
\begin{align}
\Sigma_{dd}^{<}(\omega;U)&= \Sigma_{dd}^{(1)<}(\omega;U)  +[(1-\gamma)g\omega_0]^2\nonumber\\
\times \Big [ A&(\omega-\omega_0;U) \bar f(\omega-\omega_0;U)n_B(\omega_0) \nonumber\\
+A(&\omega+\omega_0;U) \bar f(\omega+\omega_0;U) (n_B(\omega_0)+1) \Big],\label{EQUSigmalessfourier} \\[0.2cm]
\Sigma_{dd}^{(1)<}(\omega;U)&= \sum_a \Big \{ I_0(\kappa)\widetilde \Gamma_a^{(0)}(\omega+\mu)n_F(\omega+U_a) \nonumber\\
+ \sum_{s\ge 1}& I_{s}(\kappa) 2 \sinh (s \theta)\Big [ \widetilde\Gamma_a^{(0)}(\omega-s\omega_0+\mu) \nonumber\\
\times n_B(&s\omega_0) n_F(\omega-s\omega_0+U_a)+ \widetilde\Gamma_a^{(0)}(\omega+s\omega_0+\mu)\nonumber\\
 \times(n_B&(s\omega_0)+1) n_F(\omega+s\omega_0+U_a)  \Big ] \Big \}, \label{EQUSigmalessfourier1}
\end{align}
with $n_F(\omega)=(e^{\beta\omega}+1)^{-1}$, $n_B(\omega)=(e^{\beta\omega}-1)^{-1}$ and
\begin{align}
\widetilde\Gamma_a^{(0)}(\omega) & = \mathrm{e}^{-\gamma^2 g^2\coth\theta}\Gamma_a^{(0)}(\omega), \\
\Gamma_a^{(0)}(\omega) &  =2\pi  |t_a(\omega)|^2\varrho_a(\omega),\\
\varrho_a(\omega)&=\frac{1}{N}\sum_k\delta(\omega-\varepsilon_{ka}),\\
\theta&= \frac{1}{2}\beta\omega_0\;,\quad \kappa=\frac{\gamma^2 g^2}{\sinh \theta},\\
I_s(\kappa)&=\sum_{m=0}^{\infty}\frac{1}{m!(s+m)!}\left(\frac{\kappa}{2}\right)^{s+2m}.\label{EQUdefthetakappa}
\end{align}
The function $\Sigma_{dd}^{<}(\omega;U)$ describes the in-scattering of polaron-like quasiparticles at the dot~\cite{Da95_2}. It accounts for multiple-phonon emission/absorption processes at finite temperature and with finite particle densities. $\Sigma_{dd}^{>}(\omega;U)$ results from interchanging $n_B\leftrightarrow (n_B+1)$, $n_F\leftrightarrow (1-n_F)$ and $ \bar f \leftrightarrow (1- \bar f)$ in Eq.~(\ref{EQUSigmalessfourier}). Then the spectral function follows using Eqs.~(\ref{EQUdefGamma}) and (\ref{EQUafrac}). As we see from Eq.~(\ref{EQUSigmalessfourier}), for any $\gamma<1$ the functions $A$ and $ \bar f$ have to be determined self-consistently. Moreover, the renormalized dot level (\ref{EQUdefeta}) depends on the dot occupation $n_d$, which also has to fulfill a self-consistency condition:
\begin{equation}\label{EQUnselfcons}
n_d=\int_{-\infty}^{\infty}\frac{\mathrm{d}\omega}{2\pi}\; \bar f(\omega;U) A(\omega;U).
\end{equation}
To determine the variational parameter $\gamma$, we minimize the thermodynamic potential $\Omega$. We use a decoupling approximation between the electron and oscillator degrees of freedom and neglect the influence of the dot states on the leads. We consider an ensemble given by (\ref{EQUhamiltonian}), but with the EP and DL interaction coefficients being multiplied by $\lambda\in[0,1]$. Then the thermodynamic potential follows from the well-known general relations in Refs.~\onlinecite{KB62,FW71}:
\begin{align}
\Omega&= -\frac{1}{\beta}\ln (1+\mathrm{e}^{-\beta\widetilde\eta})+ \varepsilon_p(1-\gamma)^2n_d^2\nonumber\\
&+2\int_0^{1}\mathrm{d}\lambda\;\frac{1}{\lambda}\int\frac{\mathrm{d}\omega}{2\pi}\;(\omega-\widetilde\eta)\;A_\lambda(\omega;U) \bar f_\lambda(\omega;U).\label{EQUOmega1}
\end{align}
To make the integration in Eq.~(\ref{EQUOmega1}) feasible, we determine $A_\lambda$ from Eq.~(\ref{EQUafrac}) with the self-energy in the first iteration step, i.e. $\Gamma_{\lambda}^{(1)}=\lambda^2(\Sigma_{dd}^{(1)>}+\Sigma_{dd}^{(1)<})$. Correspondingly, $ \bar f^{(1)}_\lambda$ follows from Eq.~(\ref{EQUbarf}) using $\Sigma_{dd}^{(1)<}$ and $\Gamma^{(1)}$. However, $\widetilde \eta$ will be determined from the dot occupation $n_d$ resulting from the complete self-energy. The parameter $\gamma$ that minimizes the thermodynamic potential determines $\Sigma_{dd}^{\lessgtr}(\omega;U)$ and, consequently, the complete functions $ \bar f(\omega;U)$ and $A(\omega;U)$.
%
\subsection{Electron current and linear response thermopower}
The operator of the electron current from lead $a$ to the dot reads
\begin{equation}\label{EQUcurrentop}
\hat J_a=\frac{\mathrm{i}e}{\sqrt{N}} \sum_k \left [ t_{ka} \widetilde d^\dag c_{ka}^{\nag} - t_{ka}^\ast c_{ka}^\dag \widetilde d \right ],
\end{equation}
with the negative elementary charge $e$. We determine the mean value $J_a=\langle \hat J_a \rangle$ using the connection of the required expectation values to the real-time ``mixed" Green functions $g_{cd}(k,a;t_1,t_2;U)$, which are defined similar to Eqs.~(\ref{EQUdefresponseless}) and (\ref{EQUdefresponsegtr})~\cite{KLAF11}. In the following we assume identical leads and work in the wide band approximation, i.e. we set $\Gamma_a^{(0)}(\omega)=\Gamma_0=\mathrm{const}$. Then the steady-state charge current through the dot, $J=(J_L-J_R)/2$, reads 
\begin{align}
J&=\frac{e \Gamma_0}{2}\int_{-\infty}^{\infty}\frac{\mathrm{d}\omega}{2\pi}\,\,\widetilde A(\omega;U) \left[ n_F(\omega+U_L) -  n_F(\omega+U_R) \right ] , \label{EQUcurrent}
\end{align}
with the electronic spectral function $\widetilde A (\omega;U)$.
The latter is obtained in terms of the polaronic spectral function as follows~\cite{KLAF11}:
\begin{align}\label{EQUelectronic}
\widetilde A & (\omega;U)=\\[0.1cm]
&\quad\quad\mathrm{e}^{-\gamma^2 g^2\coth\theta}\Big\{ I_{0}(\kappa)A(\omega;U) + \sum_{s\ge 1} I_{s}(\kappa)2\sinh(s\theta) \nonumber \\
&\quad\times\Big( \left[n_B(s\omega_0)+ \bar f(\omega+s\omega_0;U)\right]A(\omega+ s\omega_0;U) \nonumber \\
&\quad+ \left[n_B(s\omega_0)+1- \bar f(\omega-s\omega_0;U)\right]A(\omega- s\omega_0;U)\Big ) \Big\}.\nonumber
\end{align}
Moreover, we define the differential conductance $G$ of the quantum dot system as
\begin{align}
G&=\frac{\mathrm{d}J}{\mathrm{d}\Phi}\;.
\end{align} 
In the linear response regime, we suppose the application of an infinitesimal voltage bias $\Phi=\delta\mu/e$ and temperature difference $\delta T$ between the leads. Then we can expand the current to first order in $\delta\mu$ and $\delta T$ as~\cite{EIA10}
\begin{align}\label{EQUcoefficients}
J&= L\frac{\delta\mu}{e} +X \frac{\delta T}{T}\;, 
\end{align}
where $L$ is the linear response conductance and $X$ is the thermoelectric coefficient. Both quantities follow from the linearization of the Fermi functions in Eq.~(\ref{EQUcurrent}) around the equilibrium chemical potential $\mu$ and the equilibrium temperature $T$:
\begin{align}
L&=\lim_{\delta\mu\to0} \{ eJ/\delta\mu\}\Big|_{\delta T=0}\nonumber\\
&=\frac{e^2\Gamma_0}{2}\beta \int_{-\infty}^{\infty}\frac{\mathrm{d}\omega}{2\pi}\;\widetilde A(\omega)n_F(\omega)(1-n_F(\omega)),\label{EQUconductance}
\end{align}
\begin{align}
X&=\lim_{\delta T\to0} \{ T J/\delta T\}\Big|_{\delta \mu=0}\nonumber\\
&=\frac{e\Gamma_0 }{2}\beta\int_{-\infty}^{\infty}\frac{\mathrm{d}\omega}{2\pi}\;\omega\;\widetilde A(\omega)n_F(\omega)(1-n_F(\omega)). \label{EQUthermo}
\end{align}
In (\ref{EQUconductance}) and (\ref{EQUthermo}), the electronic spectral function is calculated in equilibrium. With the help of these transport coefficients we define the linear response thermopower
\begin{align}\label{EQUthermopower}
S&=\frac{eX}{TL}\,,
\end{align}
which is a measure of the thermoelectric efficiency of the quantum dot system.
%
%
%
\subsection{Weak EP coupling limit}
The current formula (\ref{EQUcurrent}) and the expressions for the linear response coefficients in Eqs.~(\ref{EQUconductance}) and (\ref{EQUthermo}) have a simple structure, because all effects of the EP interaction are contained in the electronic spectral function $\widetilde A$. However, our approximation to the spectral function includes terms of arbitrarily high order in the EP coupling strength $g$: For $\gamma>0$, this can be seen explicitly in the summations over $s$ in Eqs.~(\ref{EQUSigmalessfourier1}) and (\ref{EQUelectronic}), which describe inelastic (quasielastic) processes involving the emission and absorption of an unequal (equal) number of phonons.
As long as $\gamma<1$, high order terms will also result from the iterative calculation of the self-consistent equation (\ref{EQUSigmalessfourier}). Via the denominator of the polaronic spectral function in Eq.~(\ref{EQUafrac}) the transport channels will be affected by a voltage dependent renormalization of the effective dot level and the real part of the self-energy. Lastly, all of these contributions are functions of the optimal parameter $\gamma_{\mathrm{min}}$, which itself will be voltage dependent. This will lead to complicated current-voltage characteristics in the numerical evaluation of Eq.~(\ref{EQUcurrent}), which are presented in the next section.

For a better understanding of the numerical results, we want to gain more insight on the different EP coupling effects and their dependence on the parameter $\gamma$. To this end, we consider the limit of small EP coupling strengths $g$ and low voltages $\Phi < 2\omega_0$. Then we can expand the self-energy and the spectral function to second order in $g$ around the noninteracting (i.e. zeroth-order) results. In doing so, we work in the wide-band approximation $\Gamma_a^{(0)}(\omega)=\Gamma_0=\mathrm{const}$ and consider low temperatures $T\ll\omega_0$, so that $n_B(\omega_0)\approx 0$. First, we set $g=0$ in Eqs.~(\ref{EQUSigmalessfourier}) and (\ref{EQUSigmalessfourier1}) and obtain the zeroth-order functions
\begin{align}
\Gamma^{(0)}(\omega)&= 2\Gamma_0 \;,\label{EQUGammazero}\\
A^{(0)}(\omega) &=  \frac{2\Gamma_0}{(\omega-\Delta+\mu)^2+\Gamma_0^2}\;, \label{EQUAzero}\\
\bar f^{(0)}(\omega;U) &= \frac{1}{2}\Big ( n_F(\omega+U_L) + n_F(\omega+U_R) \Big )\;, \label{EQUfbarzero}\\
n_d^{(0)}&=\int_{-\infty}^{\infty}\frac{\mathrm{d}\omega}{2\pi}\; \bar f^{(0)}(\omega;U) A^{(0)}(\omega)\;.\label{EQUnzero}
\end{align}
Equations~(\ref{EQUGammazero})-(\ref{EQUnzero}) are the exact solution for $g=0$ and describe a rigid quantum dot acting as a tunneling barrier between the leads.
Next, we insert $A^{(0)}$ and $\bar f^{(0)}$ for $A$ and $\bar f$ in Eq.~(\ref{EQUSigmalessfourier}), which corresponds to the first step in the self-consistent calculation. Moreover, for $T\ll\omega_0$, we expand the r.h.s. of Eq.~(\ref{EQUSigmalessfourier1}) to second order in $g$, whereby only the terms with $s=0,1$ contribute. The resulting approximation of the function $\Gamma$ can be written as
\begin{align}
\Gamma(\omega;U) &\approx \Gamma^{(0)}(\omega) +\Gamma^{(2)}(\omega;U)\;,
\end{align}
with the second order correction
\begin{align}
 \Gamma^{(2)}(\omega;U)&= -2\gamma^2 g^2 \Gamma_0 + 2\gamma^2 g^2 \Gamma_0 \Big( \bar f^{(0)}(\omega+\omega_0;U) \nonumber\\
+ &1- \bar f^{(0)}(\omega-\omega_0;U)\Big) \nonumber\\
+& [(1-\gamma)g\omega_0]^2 \Big [ A^{(0)}(\omega+\omega_0) \bar f^{(0)}(\omega+\omega_0;U) \nonumber\\
+& A^{(0)}(\omega-\omega_0) (1-\bar f^{(0)}(\omega-\omega_0;U)) \Big]\;.\label{EQUGammacorrection}
\end{align}
The second order renormalization of the dot level results from substituting $n_d^{(0)}$ for $n_d$ in Eq.~(\ref{EQUnselfcons}). Then $\widetilde\eta$ is approximated as $\widetilde\eta \approx \Delta -\mu + \widetilde\eta^{(2)}$, with
\begin{align}
\widetilde\eta^{(2)}&=-\varepsilon_p\gamma(2-\gamma)-2\varepsilon_p(1-\gamma)^2n_d^{(0)}\;.\label{EQUncorrection}
\end{align}
Consequently, we expand the polaronic spectral function in Eq.~(\ref{EQUafrac}) with respect to the second order corrections $\Gamma^{(2)}$ and $\widetilde\eta^{(2)}$ and obtain 
\begin{align}
A(\omega) &\approx A^{(0)}(\omega) + A^{(2)}(\omega;U) \;,\label{EQUpolaronicapprox}
\end{align}
with
\begin{align}
A^{(2)}(\omega;U)&=\left (\frac{A^{(0)}(\omega)}{2\Gamma_0}\right)^2\nonumber \\
& \times\Big\{\; 4\Gamma_0(\omega-\Delta+\mu) \left ( \widetilde\eta^{(2)} + \mathrm{Re} \Sigma^{(2)}_{dd}(\omega;U) \right )\nonumber \\
& + \Big ( (\omega-\Delta+\mu)^2 -\Gamma_0^2 \Big) \Gamma^{(2)}(\omega;U) \;\Big \}\;\label{EQUpolaroniccorrection}
\end{align}
and
\begin{align}
\mathrm{Re}\Sigma_{dd}^{(2)}(\omega;U)&=\mathcal{P}\int\frac{\mathrm{d}\omega^\prime}{2\pi}\frac{\Gamma^{(2)}(\omega^\prime;U)}{\omega-\omega^\prime}\label{EQUReSigamcorrection}\;.
\end{align}
Now we replace the polaronic spectral functions on the r.h.s. of Eq.~(\ref{EQUelectronic}) with the approximation in Eq.~(\ref{EQUpolaronicapprox}), and keep only terms up to second order in $g$. Then the small coupling approximation to the electronic spectral function follows as 
\begin{align}
\widetilde A (\omega;U) &\approx  A^{(0)}(\omega) -\gamma^2 g^2 A^{(0)}(\omega) + A^{(2)}(\omega;U) \nonumber\\
& + \gamma^2 g^2 \Big [ A^{(0)}(\omega+\omega_0) \bar f^{(0)}(\omega+\omega_0;U) \nonumber\\
 &+ A^{(0)}(\omega-\omega_0) (1-\bar f^{(0)}(\omega-\omega_0;U)) \Big]\;.\label{EQUelectronicapprox}
\end{align}
If we insert $\Gamma^{(2)}$ from Eq.~(\ref{EQUGammacorrection}) into Eq.~(\ref{EQUpolaroniccorrection}) and substitute the resulting expression for $A^{(2)}$ in Eq.~(\ref{EQUelectronicapprox}), the electronic spectral function can be written as the sum of five terms,
\begin{align}
\widetilde A (\omega;U) &\approx  A^{(0)}(\omega) + \widetilde A^{(2)}_{\mathrm{DL}}(\omega) + \widetilde A^{(2)}_{\Sigma}(\omega;U) \nonumber\\
&+ \widetilde A^{(2)}_{\mathrm{\eta}}(\omega;U) + \widetilde A^{(2)}_{\mathrm{inel}}(\omega;U)\;,
\end{align}
whereby $A^{(0)}$ is given in Eq.~(\ref{EQUAzero}) and we have defined
\begin{align}
\widetilde A^{(2)}_{\mathrm{DL}}(\omega) =& -\frac{\gamma^2g^2}{\Gamma_0}\Big(A^{(0)}(\omega)\Big)^2 (\omega-\Delta+\mu)^2\;,\label{EQUAFC}\\
\widetilde A^{(2)}_{\eta}(\omega;U)&=  \frac{1}{\Gamma_0}\Big(A^{(0)}(\omega)\Big)^2 (\omega-\Delta+\mu)\;  \widetilde\eta^{(2)}\;,\label{EQUAeta}\\
\widetilde A^{(2)}_{\Sigma}(\omega;U)&=  \frac{1}{\Gamma_0}\Big(A^{(0)}(\omega)\Big)^2 (\omega-\Delta+\mu)\; \mathrm{Re} \Sigma_{dd}^{(2)}(\omega;U)\;, \label{EQUAReSigma}\\
\widetilde A^{(2)}_{\mathrm{inel}}(\omega;U) &= \gamma^2 g^2 \Big [ A^{(0)}(\omega+\omega_0) \bar f^{(0)}(\omega+\omega_0;U)\nonumber\\
&+ A^{(0)}(\omega-\omega_0) (1-\bar f^{(0)}(\omega-\omega_0;U)) \Big] \nonumber\\
&+ \left(\frac{A^{(0)}(\omega)}{2\Gamma_0}\right)^2\Big ( (\omega-\Delta+\mu)^2 -\Gamma_0^2 \Big) \nonumber\\
&\times\Big\{ 2\gamma^2g^2\Gamma_0 \Big( \bar f^{(0)}(\omega+\omega_0;U) \nonumber\\
&+ 1- \bar f^{(0)}(\omega-\omega_0;U)\Big) \nonumber\\
&+ [(1-\gamma)g\omega_0]^2 \Big [ A^{(0)}(\omega+\omega_0) \bar f^{(0)}(\omega+\omega_0;U) \nonumber\\
&+ A^{(0)}(\omega-\omega_0) (1-\bar f^{(0)}(\omega-\omega_0;U)) \Big]\Big\}\;. \label{EQUAinel}
\end{align}

The function $\widetilde A^{(2)}_{\mathrm{DL}}(\omega)$ results from the second term on the r.h.s. of Eq.~(\ref{EQUelectronicapprox}) and the first term in Eq.~(\ref{EQUGammacorrection}). It accounts (to second order) for the polaronic renormalization of the DL coupling, which gives an overall reduction of the electronic density of states, apart from the resonance at $\omega=\Delta-\mu$. The terms $\widetilde A^{(2)}_{\eta}(\omega;U)$  and $\widetilde A^{(2)}_{\mathrm{\Sigma}}(\omega;U)$ represent the voltage dependent renormalization of the energy levels and contain $\gamma$ implicitly. Finally, $\widetilde A^{(2)}_{\mathrm{inel}}(\omega;U)$ denotes the inelastic contribution to the spectral function, which results from tunneling processes that involve the emission of a single phonon at the quantum dot. It includes all the terms in the electronic spectral function (\ref{EQUelectronicapprox}) that contain the functions $\bar f^{(0)}(\omega+\omega_0;U)$ and $1-\bar f^{(0)}(\omega-\omega_0;U)$ explicitly. As a consequence, it is finite only for $|\omega|>\omega_0$ and produces phononic sidebands in the dot spectrum. However, via $\mathrm{Re}\Sigma_{dd}^{(2)}$ the inelastic channels also contribute to the renormalization of the spectrum at $|\omega|<\omega_0$.
Most notably, for $\omega\to \pm\omega_0+U_a$, $\mathrm{Re}\Sigma_{dd}^{(2)}$ causes logarithmic divergences in the spectral function. If we evaluate the function $\bar f^{(0)}$ for $T\to0$ in Eq.~(\ref{EQUGammacorrection}), then $\mathrm{Re}\Sigma_{dd}^{(2)}$ follows from Eq.~(\ref{EQUReSigamcorrection}) and contains the logarithmic divergent term
\begin{align}
 (1-\gamma)^2\,\frac{g^2\omega_0^2\Gamma_0}{4\pi} & \left\{  \sum_a\frac{\ln \left ( (\omega-\omega_0+U_a)^2\right )}{ (\omega-\omega_0-\Delta+\mu)^2+\Gamma_0^2}\right.\nonumber\\
 -&\left.\sum_a\frac{ \ln \left ( (\omega+\omega_0+U_a)^2  \right )}{ (\omega+\omega_0-\Delta+\mu)^2+\Gamma_0^2} \right\} \label{EQUrealcorrection}\;.
\end{align}%
This term corresponds to the result of Entin-Wohlman {\it et al}~\cite{EIA09}, but is modified by the prefactor $(1-\gamma)^2$. Moreover, there is a new contribution to $\mathrm{Re}\Sigma_{dd}^{(2)}$, namely the term
\begin{align}
&\gamma^2 \,\frac{g^2 \Gamma_0}{4\pi} \sum_a \ln \left ( \frac{ (\omega-\omega_0+U_a)^2 }{ (\omega+\omega_0+U_a)^2 } \right ) \;.
\end{align}
For $\Phi=0$ the logarithmic divergence appearing in $\mathrm{Re}\Sigma_{dd}^{(2)}(\omega;U)$ for $\omega\to\omega_0$ has the overall prefactor
\begin{align}
\frac{g^2\Gamma_0}{4\pi}\Big(\gamma^2+\frac{(1-\gamma)^2\omega_0^2}{ (\Delta-\mu)^2+\Gamma_0^2} \Big)\;,\label{EQUdivweight}
\end{align}
so that in the adiabatic (antiadiabatic) limit $\omega_0\ll\Gamma_0$ ($\omega_0\gg\Gamma_0$), an increase in $\gamma$ raises (lowers) the overall weight of the divergences in the spectral function.

If we insert Eqs.~(\ref{EQUAFC})-(\ref{EQUAinel}) into the current formula (\ref{EQUcurrent}), we get the respective second order corrections to the noninteracting current $J^{(0)}$ and the differential conductance, i.e.
\begin{align}
J&\approx J^{(0)}+J^{(2)}_{\mathrm{DL}}+J^{(2)}_{\eta}+J^{(2)}_{\Sigma}+J^{(2)}_{\mathrm{inel}}\;,\label{EQUcurrentapprox}\\
G&\approx G^{(0)}+G^{(2)}_{\mathrm{DL}}+G^{(2)}_{\eta}+G^{(2)}_{\Sigma}+G^{(2)}_{\mathrm{inel}}\;.\label{EQUconductanceapprox}
\end{align}
For example, for $T\to0$ the second order inelastic tunneling current reads
\begin{align}
J&^{(2)}_{\mathrm{inel}}=\frac{e^2\Gamma_0^2g^2}{4\pi} \Theta(\Phi-\omega_0)\label{EQUcurrentinco}\\
&\times \left(  \int_{-U_R}^{-U_L-\omega_0}\mathrm{d}\omega
\Big \{ \;\gamma^2\frac{(\omega-\Delta+\mu)^2-\Gamma_0^2}{\left[(\omega-\Delta+\mu)^2+\Gamma_0^2 \right]^2} \right. \nonumber\\
&+\gamma^2 \frac{1}{(\omega+\omega_0-\Delta+\mu)^2+\Gamma_0^2} + (1-\gamma)^2\omega_0^2   \nonumber \\
&\times \frac{(\omega-\Delta+\mu)^2-\Gamma_0^2}{\left [ (\omega+\omega_0-\Delta+\mu)^2+\Gamma_0^2\right]\;\left [(\omega-\Delta+\mu)^2+\Gamma_0^2\right]^2} \;\Big\} \nonumber \\
&+  \int_{-U_R+\omega_0}^{-U_L}\mathrm{d}\omega
\Big \{ \;\gamma^2 \frac{(\omega-\Delta+\mu)^2-\Gamma_0^2}{\left[(\omega-\Delta+\mu)^2+\Gamma_0^2 \right]^2} \nonumber\\
&+\gamma^2\frac{1}{(\omega-\omega_0-\Delta+\mu)^2+\Gamma_0^2} + (1-\gamma)^2\omega_0^2   \nonumber \\
&\left. \times \frac{(\omega-\Delta+\mu)^2-\Gamma_0^2}{\left [ (\omega-\omega_0-\Delta+\mu)^2+\Gamma_0^2\right]\;\left [(\omega-\Delta+\mu)^2+\Gamma_0^2\right]^2} \;\Big\} \right)\nonumber\;.
\end{align}
It is finite only for $\Phi\ge\omega_0$, so that the onset of the inelastic tunneling processes will cause a jump in the differential conductance. In general, explicit analytical expressions for the second order contributions to the differential conductance can not be derived, since the optimal parameter $\gamma_{\mathrm{min}}$ is an unknown function of the voltage.
However, if we suppose that the derivative of $\gamma_{\mathrm{min}}(\Phi)$ is continuous, then for a symmetrical voltage drop $U_R=-U_L=e\Phi/2$, the jump in the differential conductance follows from Eq.~(\ref{EQUcurrentinco}) as
\begin{align}
G^{(2)}_{\mathrm{inel}}\Big|_{\Phi=\omega_0}&=\frac{e^2\Gamma_0^2g^2}{2\pi} \left ( \gamma^2 \;\dfrac{(\frac{\omega_0}{2}-\Delta+\mu)^2}{\left[ (\frac{\omega_0}{2}-\Delta+\mu)^2 + \Gamma_0^2 \right]^2} \right.\label{EQUcondinco}\\
&\hspace{-1.5cm}+\gamma^2 \;\dfrac{(-\frac{\omega_0}{2}-\Delta+\mu)^2}{\left[ (-\frac{\omega_0}{2}-\Delta+\mu)^2 + \Gamma_0^2 \right]^2} + (1-\gamma)^2\omega_0^2 \nonumber\\
&\hspace{-1.5cm}\times\left.\dfrac{(\frac{\omega_0}{2}-\Delta+\mu)^2(-\frac{\omega_0}{2}-\Delta+\mu)^2-\Gamma_0^4}
{\left[(\frac{\omega_0}{2}-\Delta+\mu)^2 + \Gamma_0^2 \right]^2\left[ (-\frac{\omega_0}{2}-\Delta+\mu)^2 + \Gamma_0^2 \right]^2} \right )\;.\nonumber
\end{align}
Again, for $\gamma\to0$ only the last term on the r.h.s. of Eq.~(\ref{EQUcondinco}) remains and coincides with the result of Entin-Wohlman {\it et al}~\cite{EIA09}. As has been discussed in Ref.~\onlinecite{EIA09}, this term is negative if the following condition is fulfilled:
\begin{align}
\Gamma_0^2>\Big| \frac{\omega_0^2}{4}-(\Delta-\mu)^2\Big|\;.\label{EQUcondition}
\end{align}
Then, at $\Phi=\omega_0$, it may cause a downward step in the differential conductance. However, the new terms $\propto \gamma^2$ in Eq.~(\ref{EQUcondinco}) are always positive. For large enough $\gamma$, they outweigh the negative contribution to (\ref{EQUcondinco}), so that the overall conductance jumps upwards, even if the condition in Eq.~(\ref{EQUcondition}) is fulfilled. 

%
%
\section{Results and Discussion}
\label{sec:numres}
In the following numerical calculations $\omega_0=1$ is fixed as the unit of energy and we set $\mu=0$ and $T=0.01$. We work in the wide band approximation, with the large bandwidth of the leads $W=60$ and $\Gamma_a^{(0)}(\omega)=\Gamma_0\Theta(\omega^2-(W/2)^2)$.
The phononic time scale is fixed by $1/\omega_0$, while the electronic time scale is given by $1/\Gamma_0$ and is used to determine which subsystem is the faster one. We will analyze the adiabatic and antiadiabatic limiting cases before considering comparable phononic and electronic time scales. In doing so, we use the ratio $\varepsilon_p/\Gamma_0$ as a measure of the EP interaction strength. 

For small to large DL coupling we calculate the polaronic spectral function $A$ and the dot occupation $n_d$ self-consistently and determine the variational parameter $\gamma_\mathrm{min}$ by numerically minimizing the thermodynamical potential $\Omega$ as a function of $\gamma$. From $A$, the electronic spectral function $\widetilde A$ as well as the linear response coefficients $L$, $X$ and the particle current $J$ follow. For finite voltages, the differential conductance $G$ is calculated numerically.

Depending on the bare dot level $\Delta$, we distinguish between the off-resonant ($\Delta\neq\varepsilon_p$) and resonant ($\Delta=\varepsilon_p$) configuration. In the latter case we find that $n_d=0.5$ is a root of (\ref{EQUnselfcons}) and we see from Eq.~(\ref{EQUdefeta}) that the renormalized dot level resonates with the equilibrium chemical potential, i.e. $\widetilde\eta=0$, for all $\gamma_{\mathrm{min}}$.
%
\subsection{Polaron induced NDC}
\label{subsec:adiab}
\begin{figure}[t]
\begin{center}
\hspace{-1cm}\includegraphics[width=0.9\linewidth]{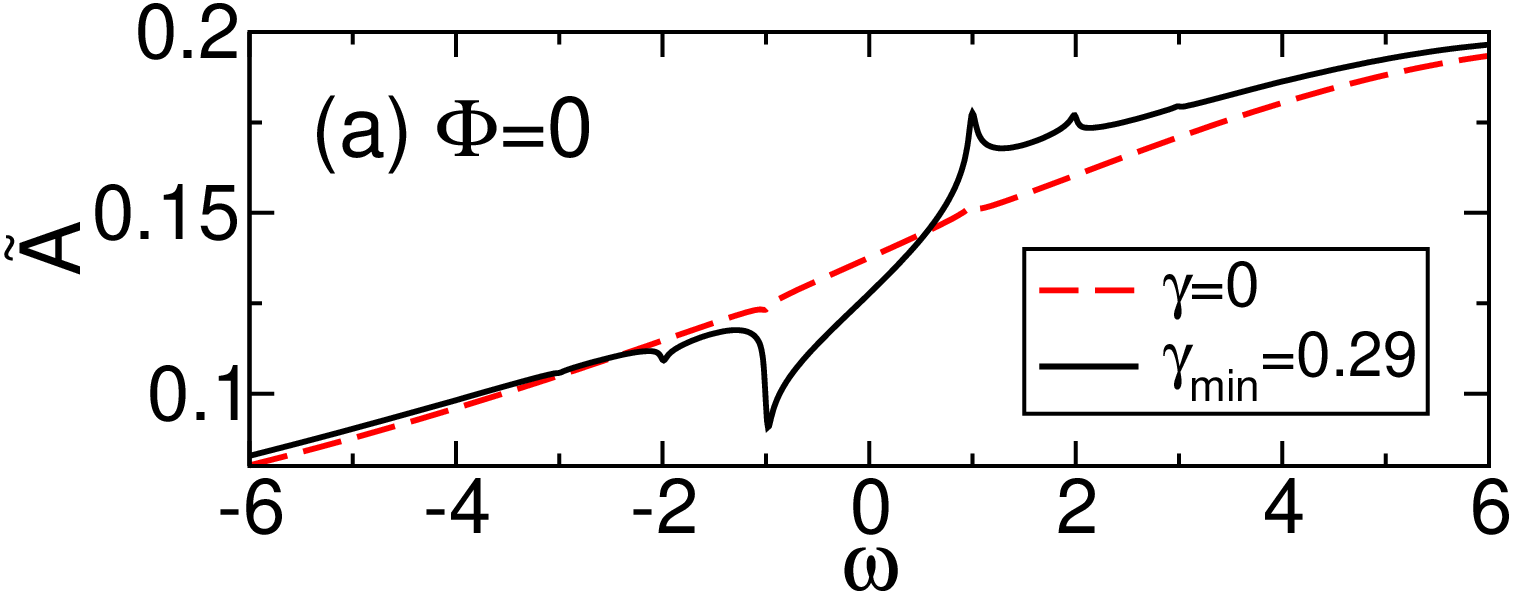}\\[0.2cm]
\hspace{-1.3cm}\includegraphics[width=0.9\linewidth]{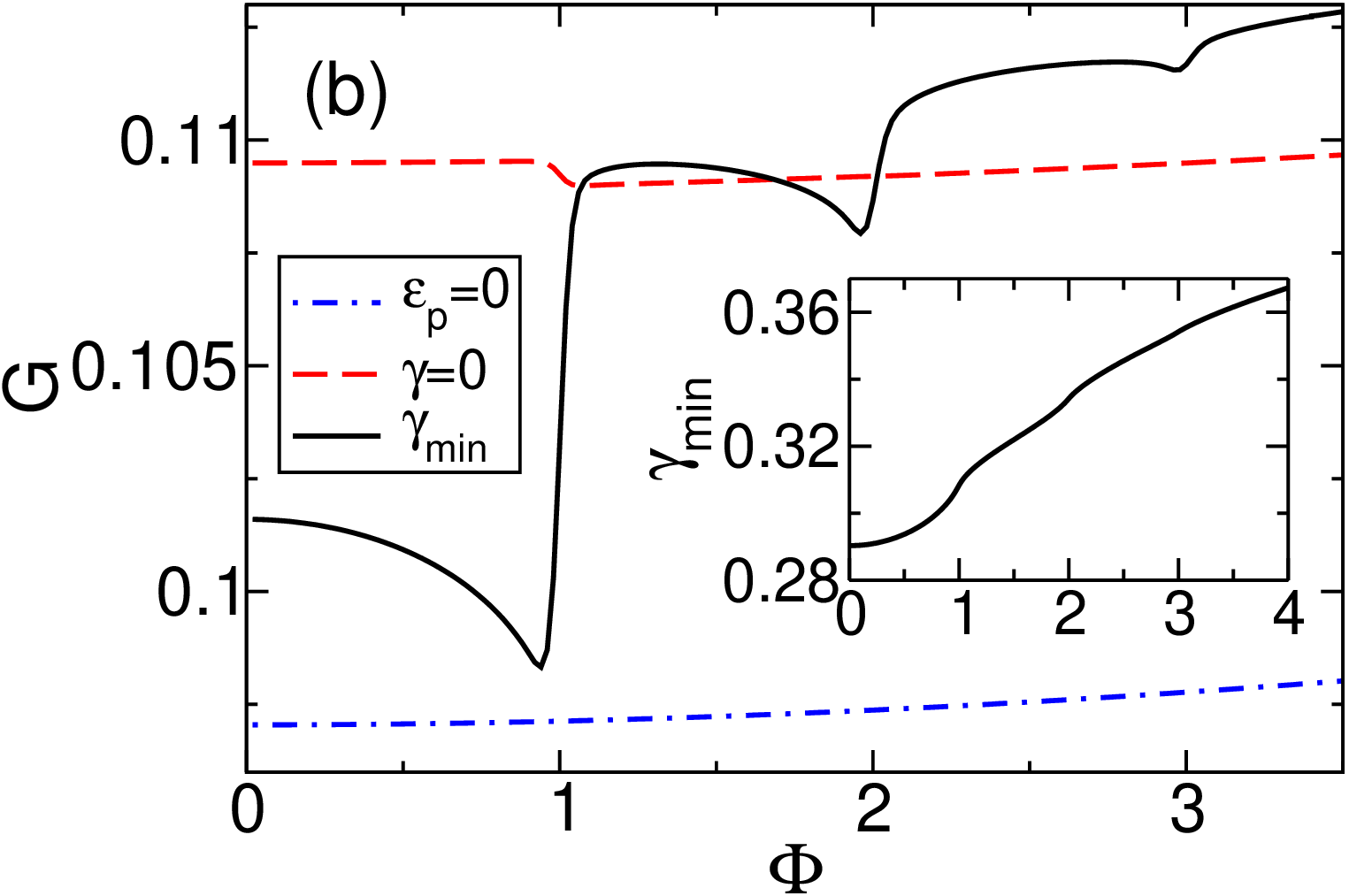}
\end{center}
\caption{For model parameters $T=0.01$, $\Gamma_0=10$, $\varepsilon_p=2$ and $\Delta=8$. Panel (a): Electronic spectral functions at $\Phi=0$ for fixed $\gamma=0$ and variationally determined parameter $\gamma_{\mathrm{min}}=0.29$. Panel (b): Differential conductance as a function of the voltage bias for $\gamma=0$ and $\gamma_{\mathrm{min}}$ in comparison to the noninteracting case $\varepsilon_p=0$. Inset: $\gamma_{\mathrm{min}}$ as a function of the voltage bias.}
\label{fig1}
\end{figure}
\begin{figure}[t]
\begin{center}
\hspace{0.0cm}\includegraphics[width=\linewidth]{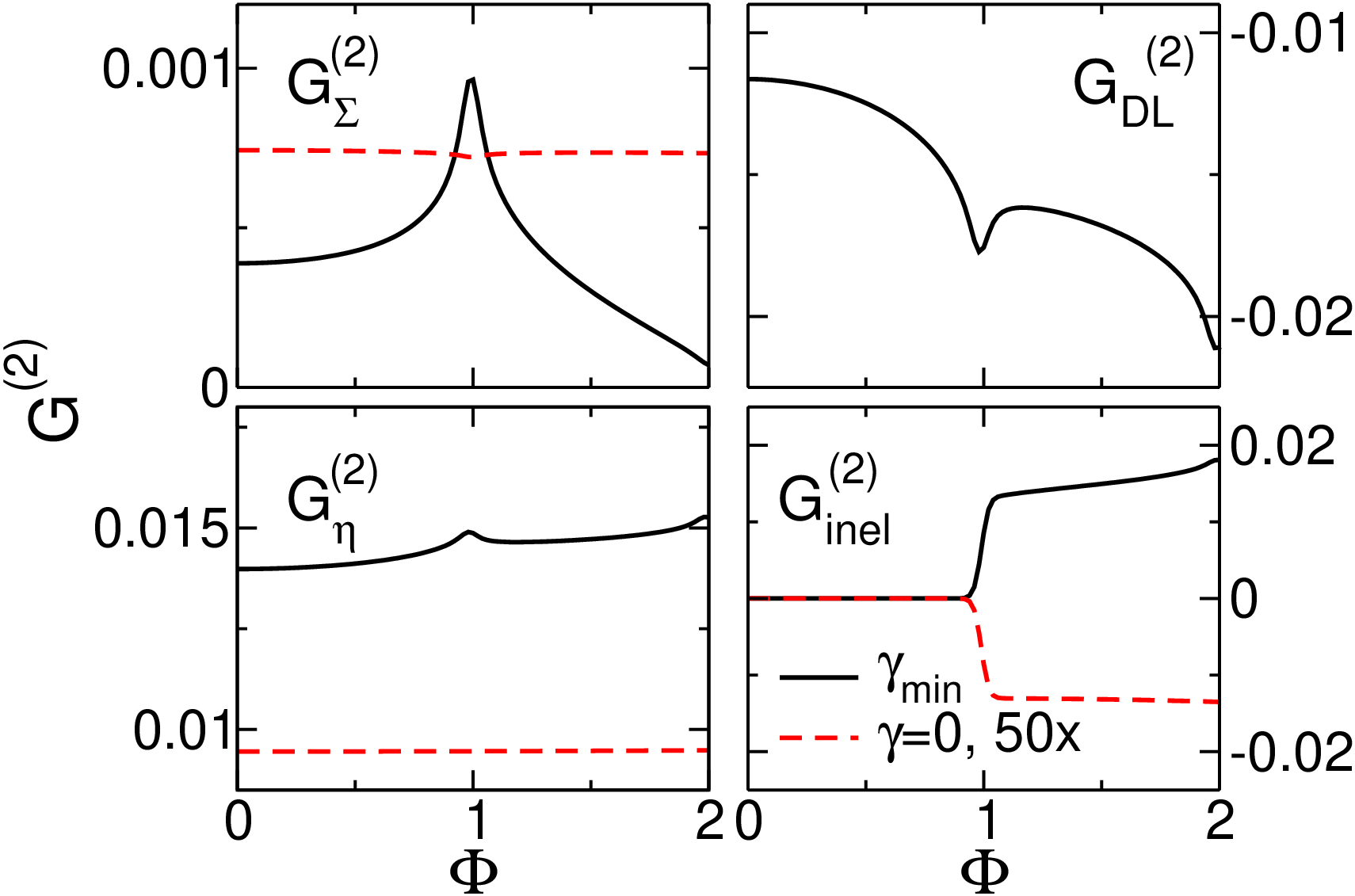}
\end{center}
\caption{For the same parameters as in Fig.~\ref{fig1}. The various second order contributions to the total differential conductance.}
\label{fig2}
\end{figure}
In their work, La Magna and Deretzis \cite{LD07} suggested the variationally determined renormalization of the dot-lead coupling as a possible mechanism for the observed nonlinear behavior of the differential conductance. We investigate whether this remains true within our approximation, which, in contrast to the effective electron model in Ref.~\onlinecite{LD07}, accounts vibrational features in the electronic spectral function to all orders in the EP coupling.

First we consider the adiabatic regime for weak EP coupling by setting $\Gamma_0=10$ and $\varepsilon_p=2$. We vary the voltage bias $0<\Phi<4$ and determine the differential conductance $G$. In doing so, we choose an off-resonant configuration with $\Delta=8$ fixed, so that the dot occupation is small and remains nearly constant during our calculations: $n_d\approx 0.3$. 

As a starting point, Fig.~\ref{fig1}(a) displays the electronic spectral function at $\Phi=0$ for the variationally determined parameter $\gamma_{\mathrm{min}}$ (black line) and compares it to the result of a calculation where we kept $\gamma=0$ fixed instead of determining $\gamma_{\mathrm{min}}$ variationally. In general, due to the large DL coupling parameter $\Gamma_0$, the electronic spectral function consists of a single wide band. For finite EP coupling, vibrational features arise at $\omega=\pm\omega_0$. These features can be attributed to logarithmic divergences in $\mathrm{Re}\Sigma_{dd}$, as the second order approximation in Eq.~(\ref{EQUrealcorrection}) suggests. While they are hardly noticeable for $\gamma=0$, the weight of the logarithmic divergences increases strongly in the variational calculation, which yields the optimal parameter $\gamma_\mathrm{min}=0.29$. This observation agrees with our discussion in the previous section: For the parameters used, Eq.~(\ref{EQUdivweight}) predicts an increase in the weight of the logarithmic contributions by a factor of about 15 with respect to the $\gamma=0$ case. Note however, that any divergences in the spectrum will be smeared out in our results due to the low but finite temperature and a numerical constraint: We evaluate the self-energy slightly above the real $\omega$ axis to prevent the unphysical loss of spectral weight. 

In Fig.~\ref{fig1}(b), the black line presents our result for the total differential conductance $G$ as a function of the voltage, with the inset showing the optimal parameter $\gamma_{\mathrm{min}}$. We compare the variational calculation to the cases $\gamma=0$ and $\varepsilon_p=0$. 
For a better understanding of the results in Fig.~\ref{fig1}(b), the four panels in Fig.~\ref{fig2} show the various second order contributions to the total differential conductance.
From Fig.~\ref{fig1}(b) it follows that for finite EP coupling the overall conductance grows with respect to the noninteracting case. Since we are considering the off-resonant regime, this can mainly be attributed to the lowering of the effective dot level. Accordingly, for $\gamma=0$, we see in Fig.~\ref{fig2} that the function $G^{(2)}_\eta$ accounts for almost all the increase in the conductance. For finite $\gamma_{\mathrm{min}}$ the effective dot level is lowered even further, but the positive contribution $G^{(2)}_\eta$ is nearly compensated by the polaronic renormalization of the DL coupling, which is shown in the upper right panel of Fig.~\ref{fig2}. With growing voltage, the optimal parameter $\gamma_{\mathrm{min}}$ increases. As the renormalization of the DL coupling grows stronger, a pronounced dip forms in the differential conductance. This mechanism is crucial for the interpretation of our calculations, as we will see below.

At $\Phi=1$, phonon emission by incident electrons becomes possible and opens up an inelastic tunneling channel. In the case $\gamma=0$, we find a small downward step in the conductance signal, since with $\Gamma_0=10$ and $\Delta=8$, the condition in Eq.~(\ref{EQUcondition}) is fulfilled. As we discussed in the previous section, for finite $\gamma_{\mathrm{min}}$ the first two terms on the r.h.s. of Eq.~(\ref{EQUcondinco}) can outweigh the third, negative term. Accordingly, our numerics show a relatively large upward step in the differential conductance (note the different scaling factors in the lower right panel of Fig.~\ref{fig2}).

%
%
\begin{figure}[t]
\begin{center}
    	\includegraphics[width=0.8\linewidth]{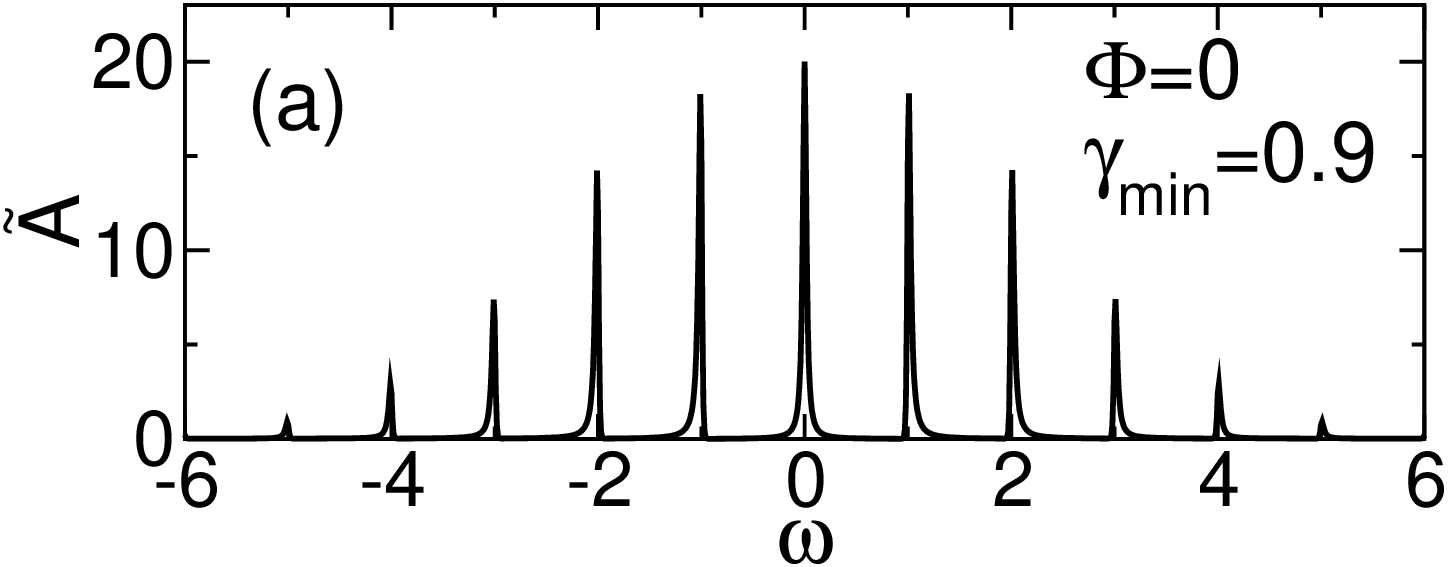}\\[0.2cm]
	\hspace{-0.4cm}\includegraphics[width=0.87\linewidth]{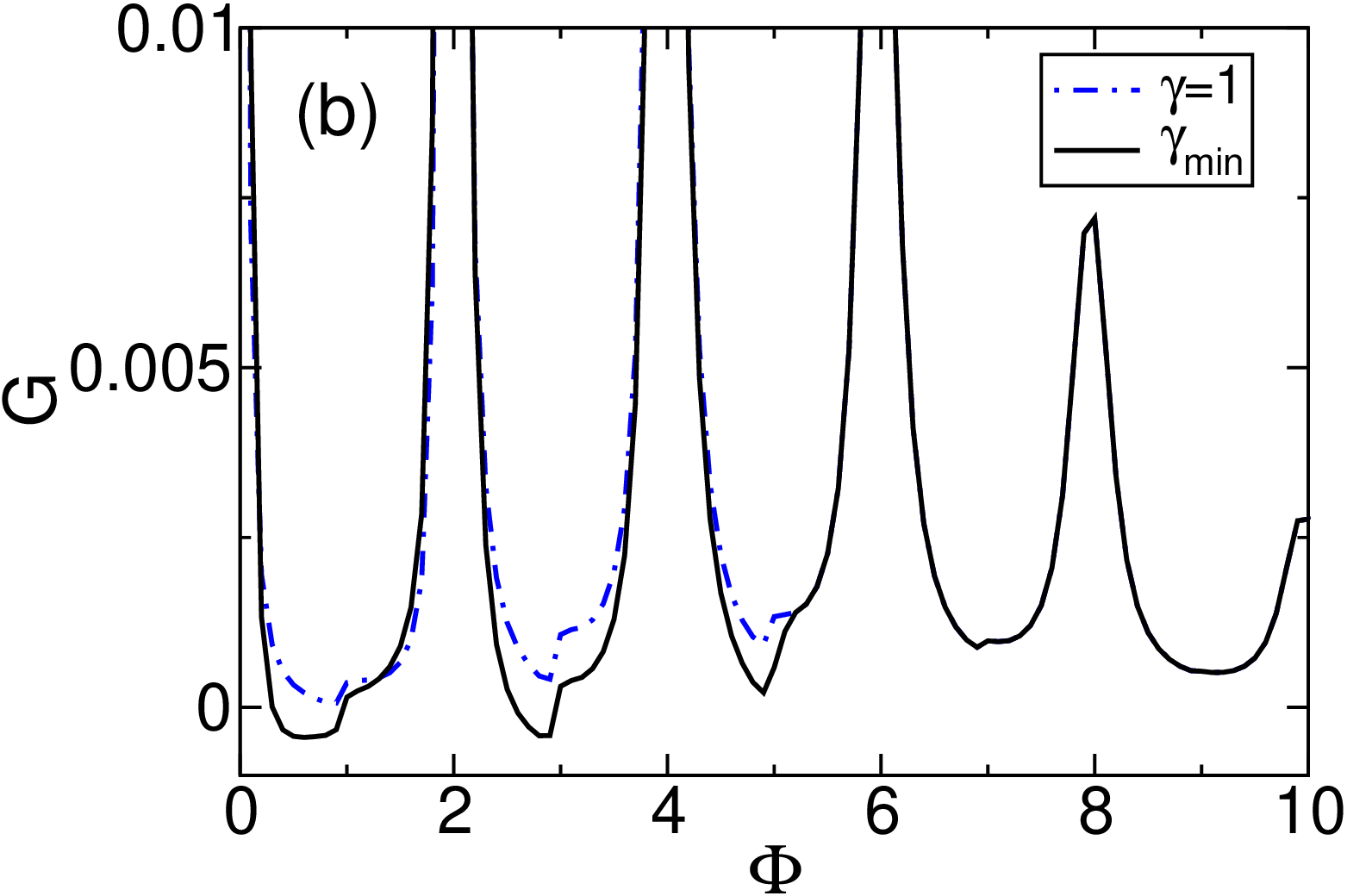}
\end{center}
\caption{For model parameters $T=0.01$, $\Gamma_0=0.1$, $\varepsilon_p=2$ and $\Delta=2$. Panel (a): Electronic spectral function of the variational calculation for $\Phi=0$ and $\gamma_\mathrm{min}=0.9$. Panel (b): Differential conductance as a function of the voltage bias, compared to the result with fixed $\gamma=1$.
}
\label{fig3}
\end{figure}
Next we investigate the polaronic renormalization in the antiadiabatic limit ($\Gamma_0=0.1$) with strong EP coupling ($\varepsilon_p=2$). We choose the resonant configuration $\Delta=\varepsilon_p$. 
For these parameters, we expect the formation of a polaron-like transient state at the quantum dot. This is confirmed by the electronic spectral function in Fig.~\ref{fig3}(a), which features several narrow phononic bands. In the low-voltage region we find $\gamma_{\mathrm{min}}\approx0.9$, i.e. the weight of the variational polaron state is smaller than predicted by the complete Lang-Firsov transformation.

Figure \ref{fig3}(b) compares the differential conductance as a function of the voltage bias for fixed $\gamma=1$ and the optimal $\gamma_{\mathrm{min}}$.
Just as in the adiabatic regime considered above, we notice small steps in the conductance at $\Phi=1,3,5$ that signal the onset of inelastic transport. In addition, a second kind of vibrational feature can be found: pronounced conductance peaks arise whenever the voltage equals multiple integers of $2\omega_0$. Here, resonant transport takes place through the polaronic side bands in $\widetilde A$.
For $\gamma=1$ the differential conductance stays strictly positive, but approaches zero between these well separated peaks. As seen for the adiabatic case, in the full calculation the polaronic renormalization grows stronger with increasing voltage bias. As a consequence, in the low voltage region the differential conductance becomes negative between the resonance peaks. Note however, that at $\Phi=1$ and $\Phi=3$ the positive nonresonant conductance steps, although carrying little weight, render the differential conductance positive again.

%
%
\begin{figure}[t]
\begin{center}
	\hspace{-1.2cm}\includegraphics[width=0.8\linewidth]{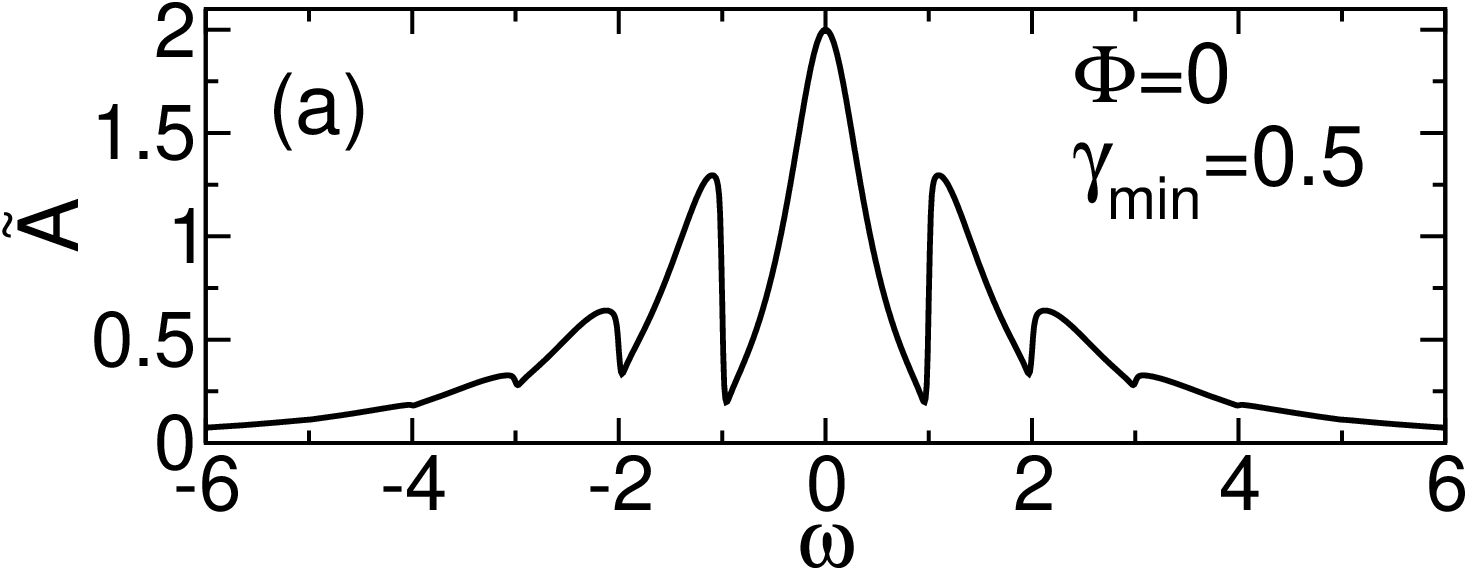}\\[0.2cm]
    	\hspace{-1.3cm}\includegraphics[width=0.82\linewidth]{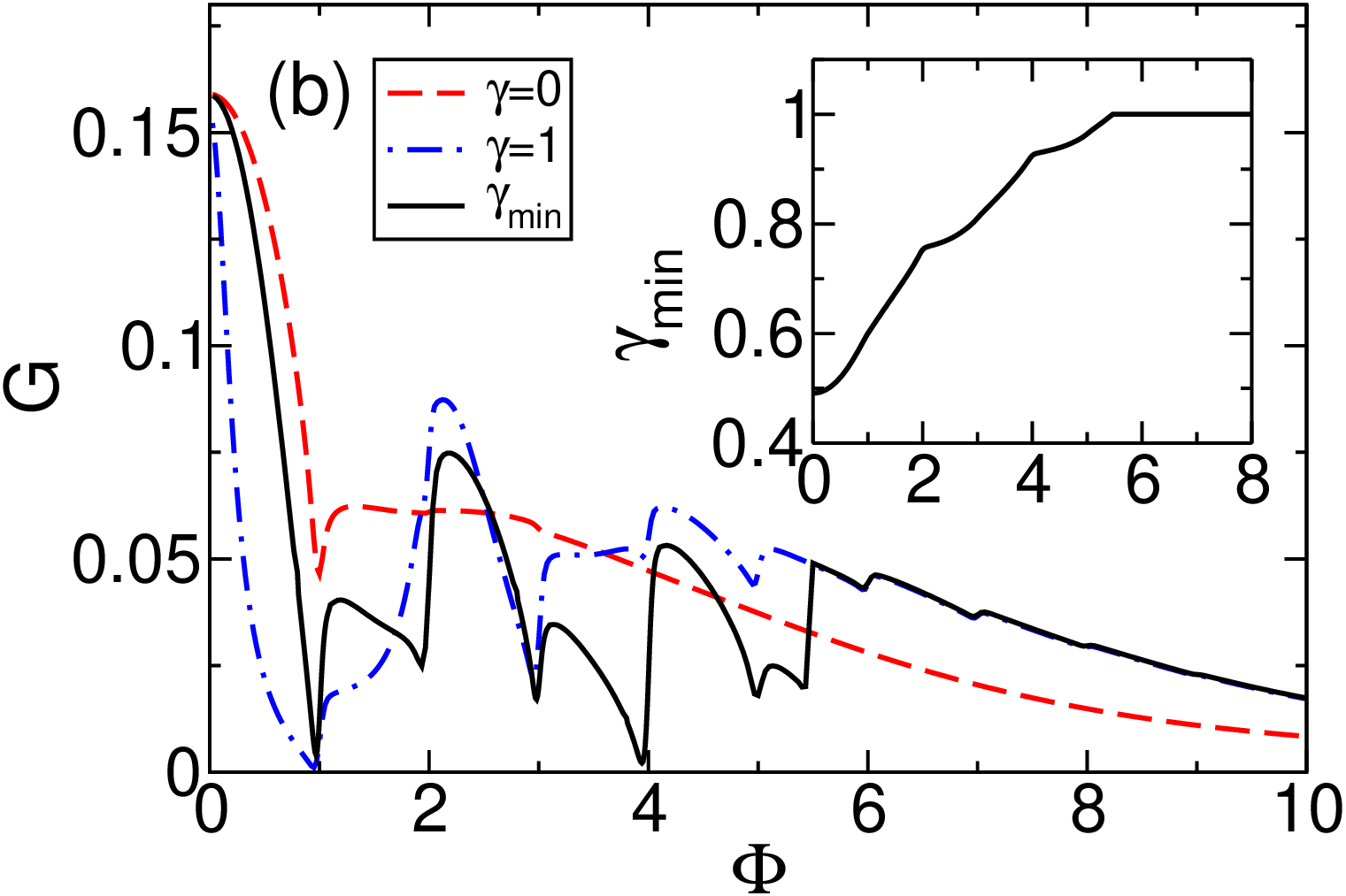}
\caption{For model parameters $T=0.01$, $\Gamma_0=1$, $\Delta=2$ and $\varepsilon_p=2$. Panel (a): Electronic spectral function in the variational calculation for $\Phi=0$ and $\gamma_\mathrm{min}=0.5$. Panel (b): Differential conductance as a function of the voltage for the variational calculation ($\gamma_{\mathrm{min}}$), compared to the results with fixed $\gamma=0$ and $\gamma=1$. Inset: Optimized variational parameter as a function of the voltage. }
\label{fig4}
\end{center}
\end{figure}
Thanks to our variational approach, we are able to investigate the interesting regime of comparable electronic and phononic energies. To this end, we set $\Gamma_0=1$ and consider intermediate EP coupling $\varepsilon_p=2$. As before, we examine the resonant, electron-hole-symmetric situation with $\Delta=2$. Fig~\ref{fig4}(a) shows the electronic spectral function at zero voltage, where the variational calculation yields $\gamma_{\mathrm{min}}\approx0.5$. Due to comparable electronic and phononic time scales, the width of the few phononic side bands is of the order of their spacing.

In Fig.~\ref{fig4}(b) we compare the conductance signal of the variational calculation to both, the $\gamma=0$ and $\gamma=1$ cases. In the low voltage regime, we have $\gamma_\mathrm{min}\gtrsim 0.5$ and the DL coupling is moderately renormalized. As the voltage grows, the variational parameter steadily increases and, as can be seen in the inset of Fig~\ref{fig4}(b), the polaron effect strengthens whenever a new resonant inelastic channel is accessible.  
The vibrational features in the conductance signal are heavily modulated by the voltage dependent polaronic renormalization: In contrast to the cases with fixed $\gamma$, there is no clear distinction between resonant peaks at $\Phi=2,4$ and off-resonant steps at $\Phi=1,3,5$, since the latter become peaks, too. Due to the comparable phononic and electronic time scales, both kinds of vibrational features have nearly the same spectral weight. Moreover, the differential conductance approaches zero between the broad conductance peaks, but no NDC is observed.

To sum up, the polaron formation involves the redistribution of spectral weight in the local density of states and, most importantly, the renormalization of the effective DL coupling. For strong EP interaction, it is indeed a possible mechanism for NDC. Yet, for small to intermediate coupling, the NDC is suppressed when multi-phonon transport processes are taken into account.
%
%
%
%
\subsection{Effective dot level}
\label{subsec:sticking}
\begin{figure}[t]
\hspace{-0.7cm}\includegraphics[width=0.82\linewidth]{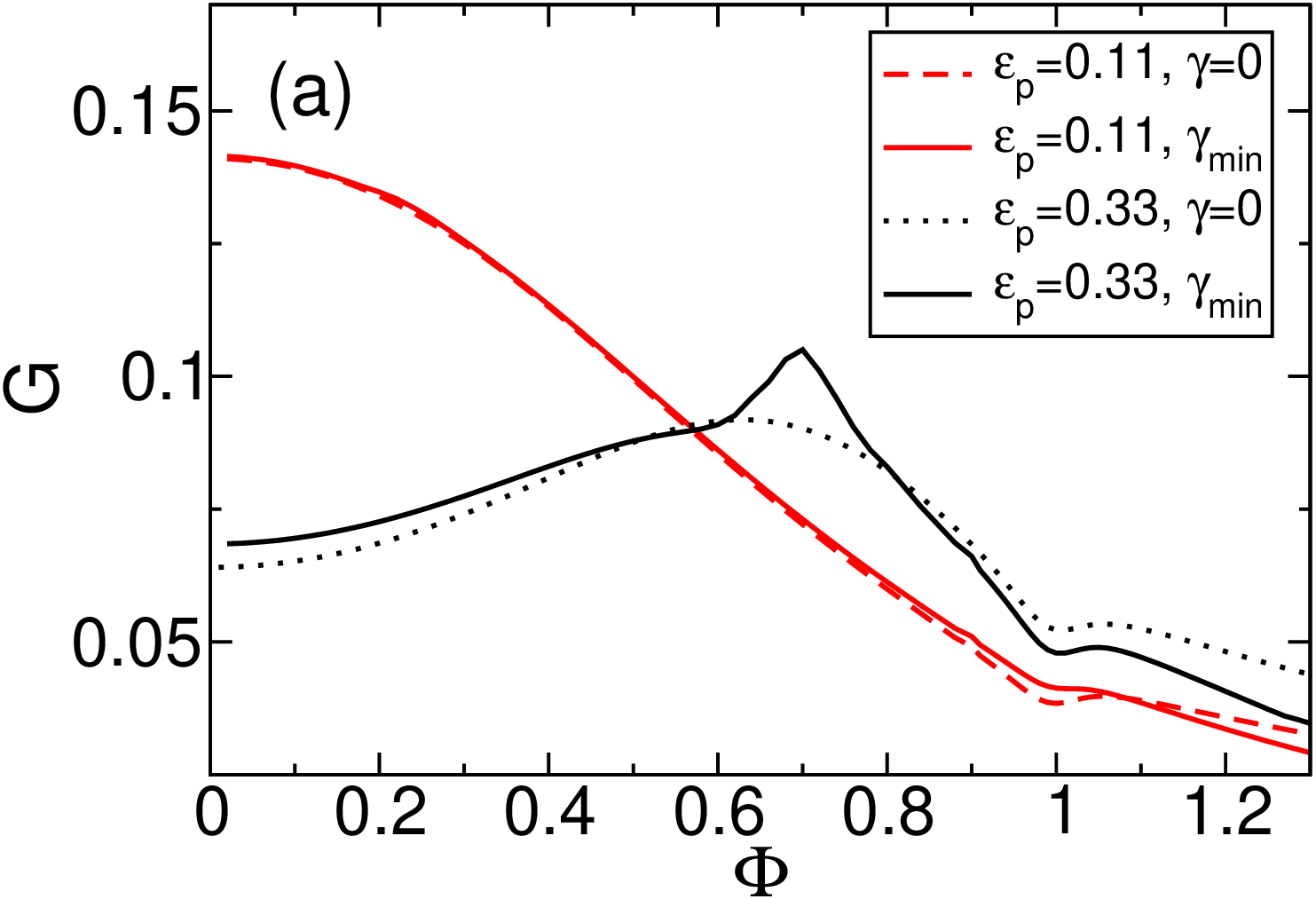}\\[0.2cm]
\includegraphics[width=.93\linewidth]{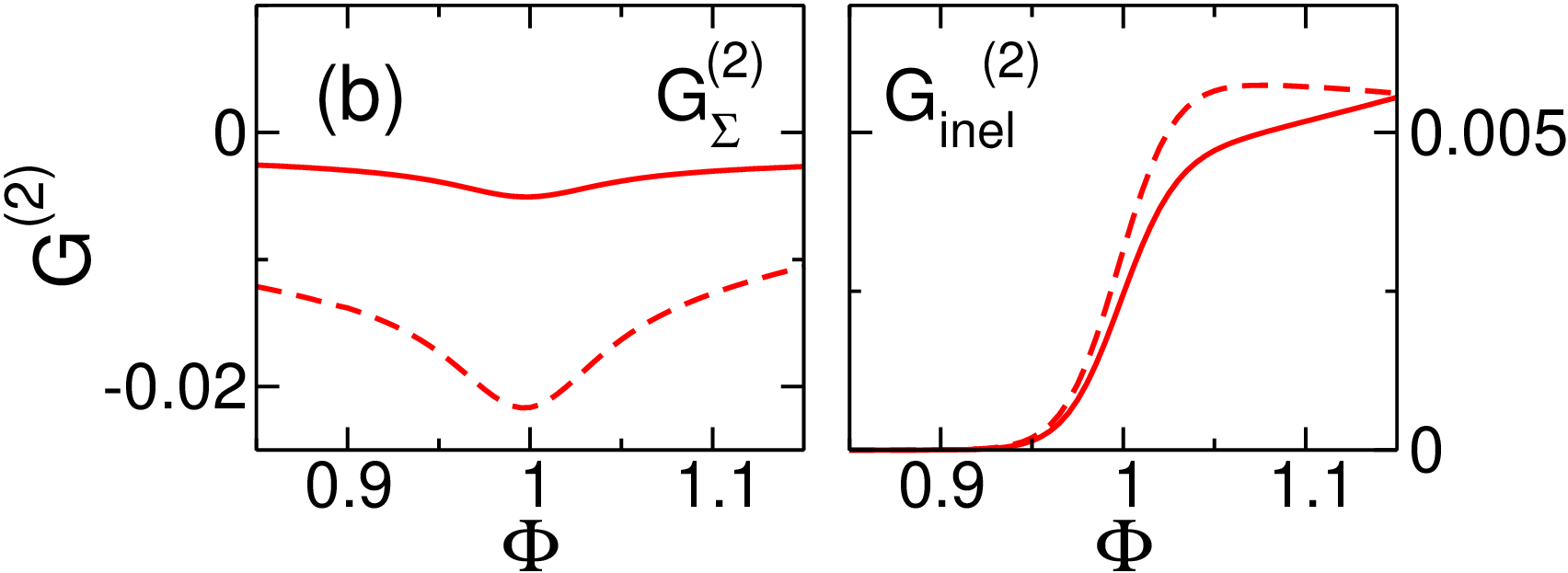}\\[0.2cm]
\hspace{-0.75cm}\includegraphics[width=0.82\linewidth]{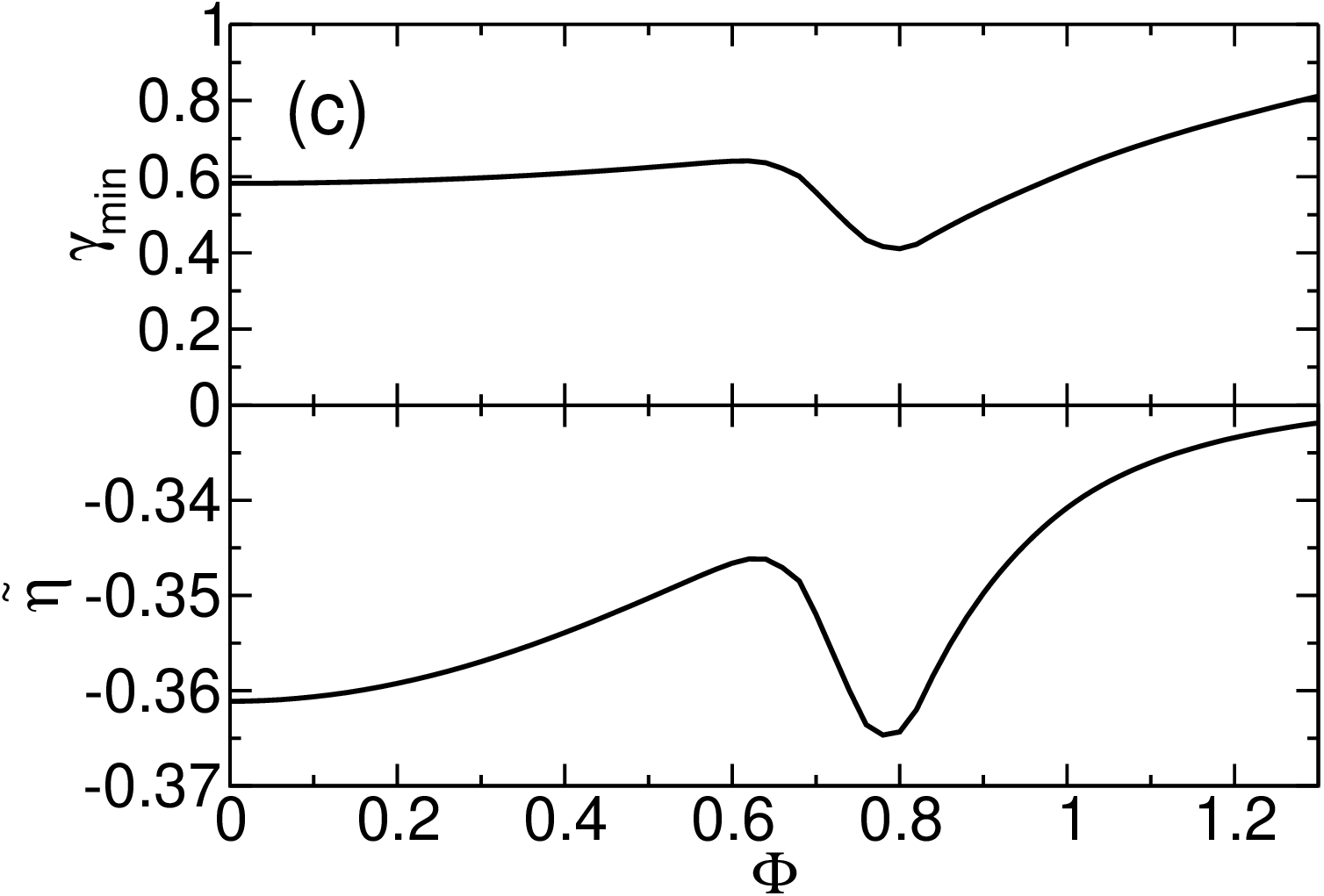}
\caption{For $T=0.01$, $\Gamma_0=0.33$ and $\Delta=0$. Panel (a): Differential conductance as a function of the voltage for the variational calculation ($\gamma_{\mathrm{min}}$), compared to the results with fixed $\gamma=0$. Panel (b): Second order contributions to the differential conductance for $\varepsilon_p=0.11$. Panel (c): Variationally determined parameter $\gamma_\mathrm{min}$ and renormalized dot-level $\widetilde\eta$ as functions of the voltage for $\varepsilon_p=0.33$.}
 \label{fig5}
\end{figure}
In the following we present another interesting consequence of the variational polaronic renormalization, which concerns the effective dot level.

We choose a slightly off-resonant configuration with $\Delta=0$ and $\varepsilon_p>0$, so that in contrast to the above calculations, the effective dot level is not pinned to the equilibrium chemical potential. We decrease the bare DL coupling slightly ($\Gamma_0=0.33$) and consider weak to intermediate EP coupling strengths. Figure~\ref{fig5}(a) compares the differential conductance for $\gamma=0$ to the variational calculation. For weak EP coupling, Fig.~\ref{fig5}(b) shows the second order approximations $G^{(2)}_{\Sigma}$ and $G^{(2)}_\mathrm{inel}$. Fig.~\ref{fig5}(c) finally presents the optimal parameter $\gamma_\mathrm{min}$ and the effective dot level $\widetilde\eta$ as functions of the voltage.

With $\varepsilon_p=0.11$, the system parameters correspond to the case of high zeroth order transmission presented in Fig.~5(a) in the work of Entin-Wohlman {\it et al}~\cite{EIA09}.
Our result for $\gamma=0$ is in good agreement with Ref.~\onlinecite{EIA09}. The conductance maximum lies near $\Phi=0$ and we find a small conductance dip at $\Phi=1$, which is caused by the logarithmic divergence in $\mathrm{Re}\Sigma_{dd}$. 
In the full calculation for $\varepsilon_p=0.11$, we find $\gamma_{\mathrm{min}}=0.75$ at $\Phi=1$. Here the dip in the total conductance vanishes. The second order approximation in the left panel of Fig.~\ref{fig5}(b) suggests that this is mainly due to a reduction of the weight of the logarithmic divergence in $\mathrm{Re}\Sigma_{dd}$. From Fig.~\ref{fig5}(b) we can also see that the jump in $G_{\mathrm{inel}}^{(2)}$ is positive for both $\gamma=0$ and $\gamma_\mathrm{min}$, since the condition (\ref{EQUcondition}) is not fulfilled for the given parameters. Moreover, in the variational calculation the height of the conductance jump is reduced with respect to the $\gamma=0$ result, which can be confirmed using Eq.~(\ref{EQUcondinco}).

If we increase the EP coupling to $\varepsilon_p=0.33$, the dip in the total conductance reappears. But most importantly, instead of a broad conductance resonance, we find a peak-like feature at $\Phi=0.7$. As we see from Fig.~\ref{fig5}(c), with increasing voltage $\widetilde\eta$ shifts upwards until at $\Phi=0.62$ it approaches the chemical potential of a lead.  For $0.65<\Phi<0.8$, the variational parameter decreases in such a way that $\widetilde\eta$ stays in resonance with the lead chemical potential. The decrease in $\gamma_\mathrm{min}$ reduces the renormalization of the DL coupling. Thereby, the system maximizes the resonant tunneling current with respect to the $\gamma=0$ case and a new peak-like conductance feature is observed in Fig.~\ref{fig5}(a). This ``sticking" of the effective dot level to the lead chemical potentials is the second main result of our variational calculations.
 
\begin{figure}[t]
\begin{center}
	\hspace{-0.7cm}\includegraphics[width=0.8\linewidth]{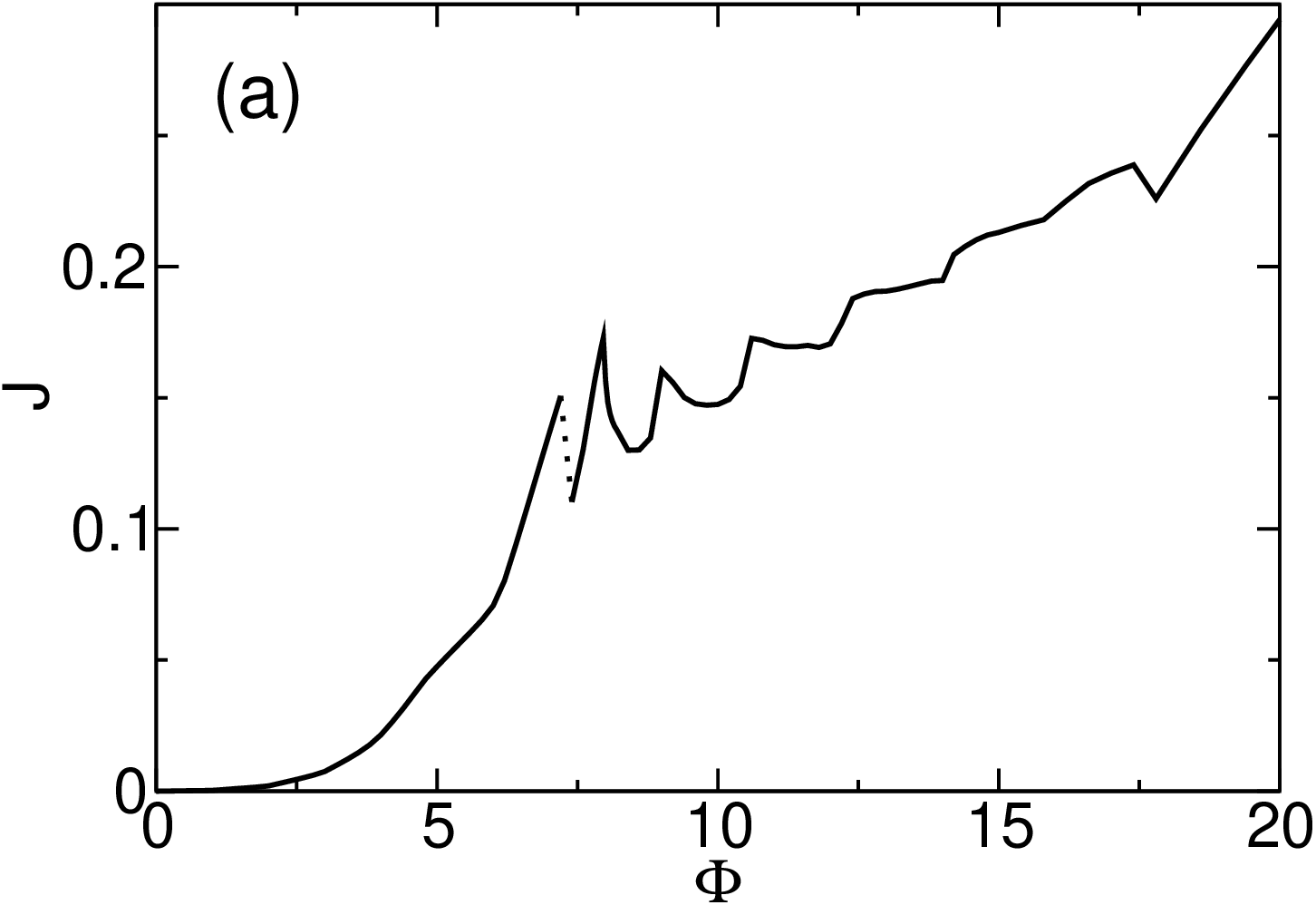}\\[0.2cm]
    	\hspace{-0.7cm}\includegraphics[width=0.8\linewidth]{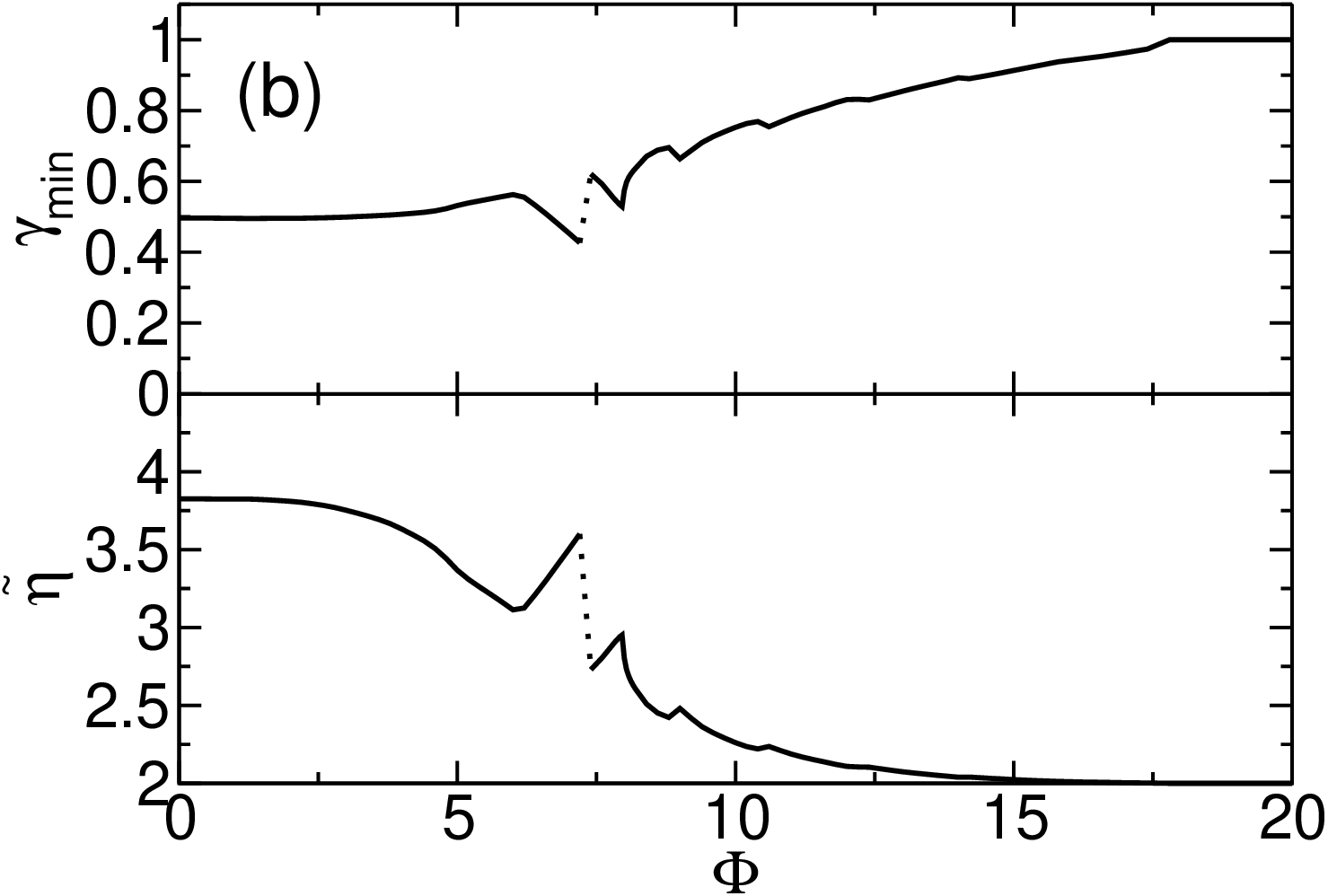}
\caption{For model parameters $T=0.01$, $\Gamma_0=1$, $\Delta=10$ and $\varepsilon_p=8$. Panel (a): Electron current as a function of the voltage. Panel (b): Variationally determined parameter $\gamma_\mathrm{min}$ and renormalized dot-level $\widetilde\eta$ as functions of the voltage.
}
\label{fig6}
\end{center}
\end{figure}

Now we consider the off-resonant scenario $\Delta=10$ for strong EP coupling $\varepsilon_p=8$. The results are presented in Fig.~\ref{fig6} (note that Fig.~\ref{fig6}(a) shows the total current). 
As expected, the effective dot level $\widetilde\eta$ sinks notably with growing voltage, until at $\Phi=6.2$ it begins to grow linearly, following the upper lead chemical potential. Again, the differential conductance grows considerably. In contrast to the intermediate EP coupling case, $\gamma_{\mathrm{min}}$ jumps from $0.4$ to $0.6$ when the system switches between two local minima in the thermodynamic potential. The resulting discontinuities in $\widetilde\eta$ and $\widetilde\Gamma_0$ cause a noticeable drop in the total current. As the voltage grows further, $\gamma_{\mathrm{min}}$ decreases again. Now the first phonon side band at $\widetilde\eta+\omega_0$ sticks to the lead chemical potential and the conductance grows once more. Similar behavior, involving the second and third side bands, is found at $\Phi\approx9$ and $\Phi\approx10.5$, respectively, until $\gamma_{\mathrm{min}} = 1$ in the high-voltage limit. Moreover, due to the strong EP coupling, the upward steps in the current are followed by regions with NDC.
\subsection{Thermopower}
\label{subsec:thermo}
\begin{figure}[t]
\begin{center}
	\hspace{-0.0cm}\includegraphics[width=0.85\linewidth]{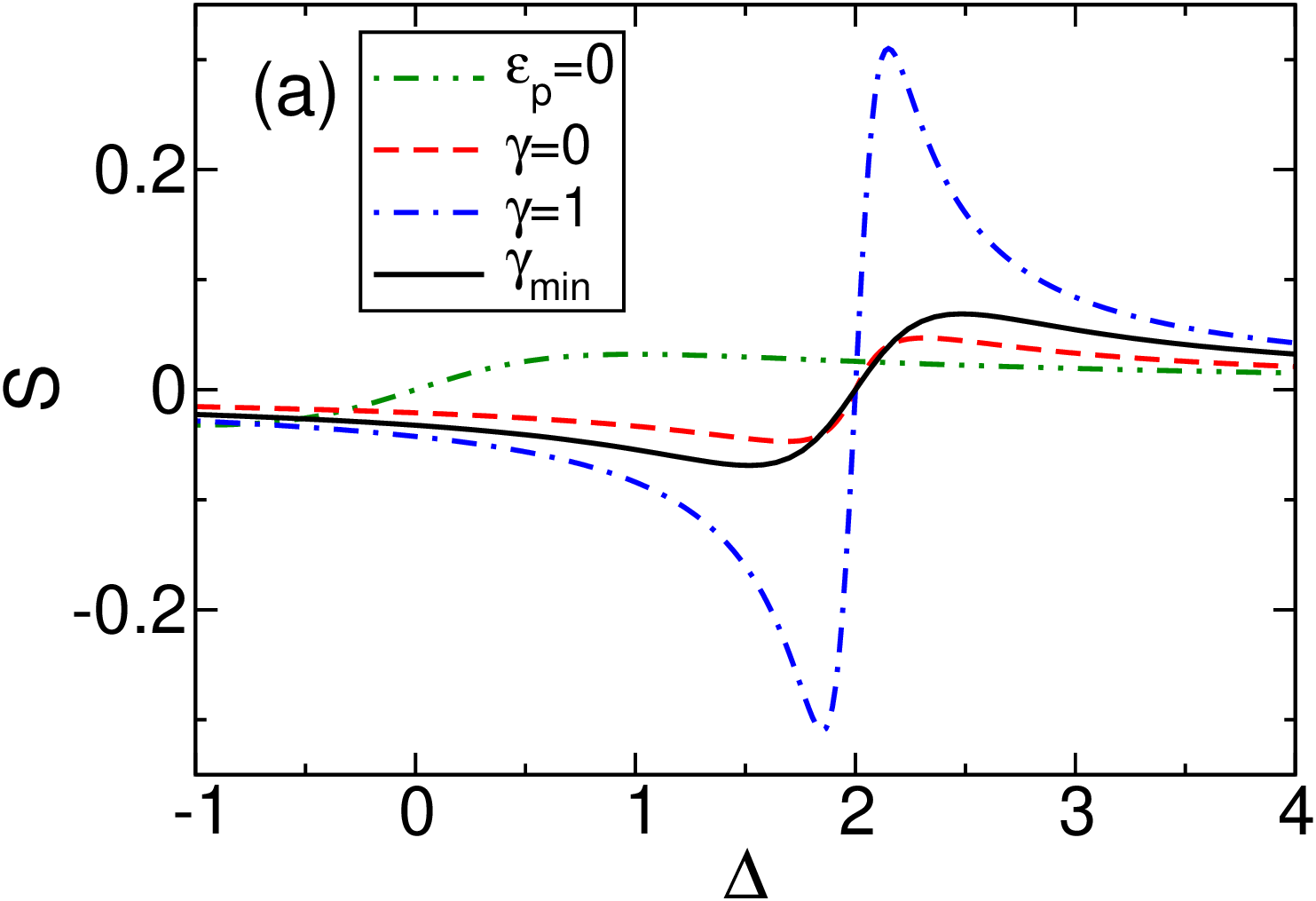}\\[0.2cm]
    	\hspace{-0.5cm}\includegraphics[width=0.9\linewidth]{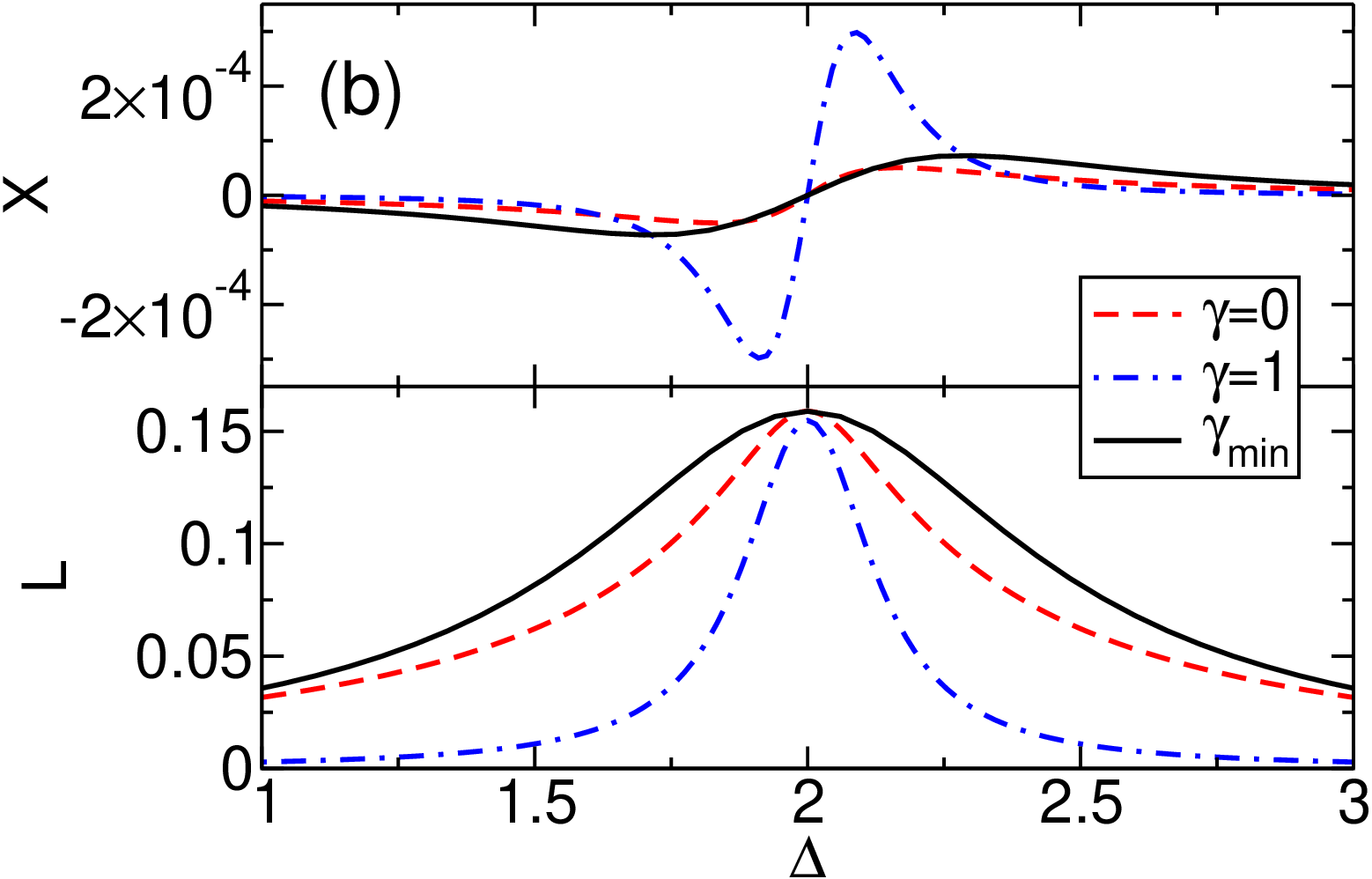}\\[0.2cm]
	\hspace{0.2cm}\includegraphics[width=0.8\linewidth]{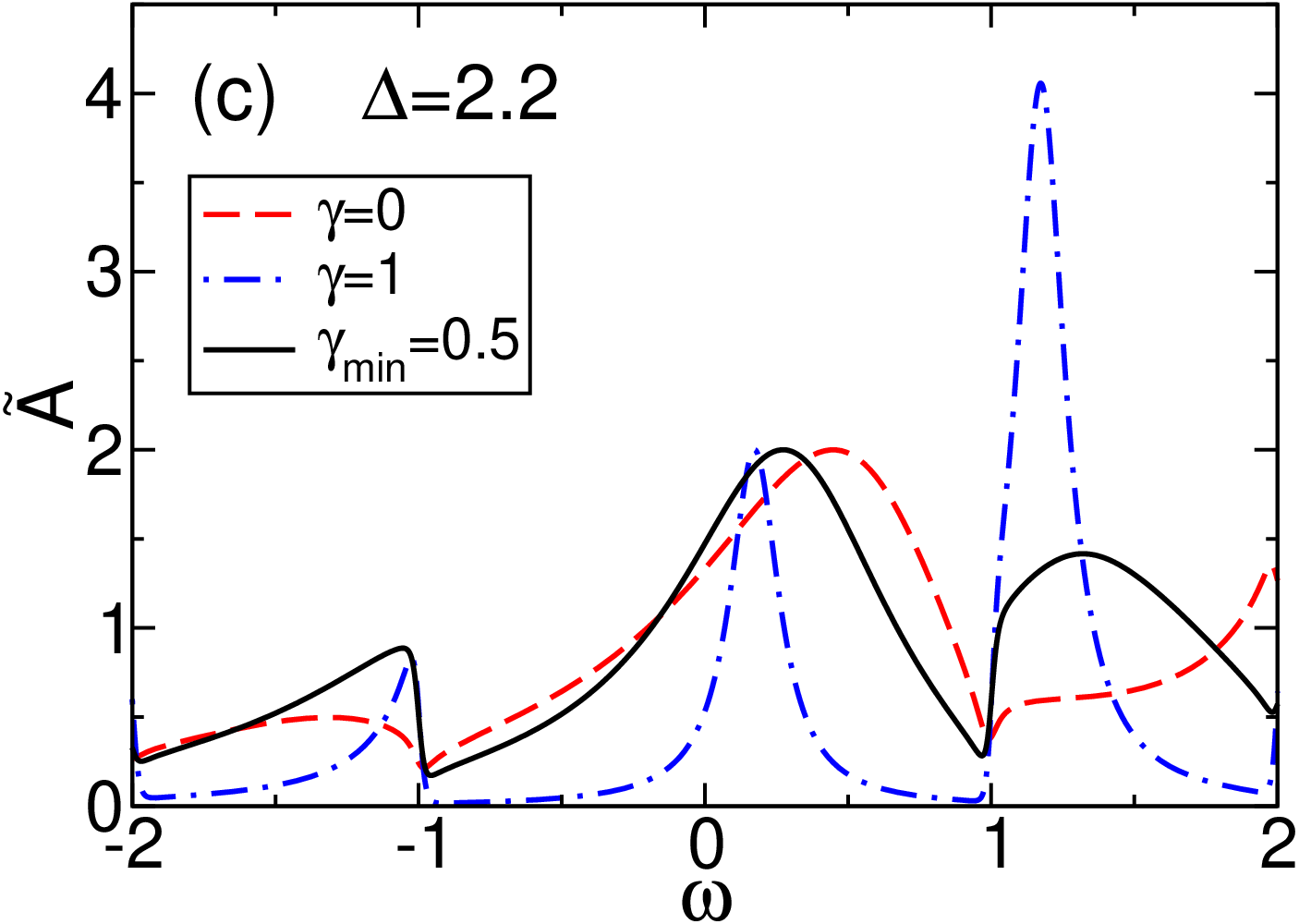}
\caption{For model parameters $T=0.01$, $\Gamma_0=1$, $\varepsilon_p=2$ and $\Phi=0$. Panel (a): Thermopower as a function of the bare dot level in the noninteracting ($\varepsilon_p=0$) and interacting system. Panel (b): Thermoelectric response $X$ and linear conductance $L$ as functions of the bare dot level. Panel (c): Electronic spectral functions at $\Delta=2.2$ for $\gamma=0$, $1$ and $\gamma_\mathrm{min}=0.5$.
}
\label{fig7}
\end{center}
\end{figure}
Finally, we investigate the thermoelectric response of the molecular junction in the physically most interesting regime of intermediate DL coupling. Setting $\Gamma_0=1$, we consider the equilibrium situation $\Phi=0$. For $\varepsilon_p=2$, we compare the variational calculation to the cases with fixed $\gamma=0,1$ and to the noninteracting system $\varepsilon_p=0$. Fig.~\ref{fig7}(a) shows the linear response thermopower $S$ as a function of the bare dot level, while Fig.~\ref{fig7}(b) presents the thermoelectric coefficient and the linear response conductance.

In general, $S$ features two resonances of opposite sign. For $\varepsilon_p=2$ they are located at $\Delta\approx\varepsilon_p\pm0.2$. In the small polaron limit $\gamma=1$ our calculation predicts a substantial increase in the maximum thermopower with respect to all the other cases. This can be explained with the help of the respective electronic spectral functions plotted in Fig.~\ref{fig7}(c) for $\Delta=2.2$. In the case $\gamma=0$, the spectral function features a broad band around the Fermi edge at $\omega=0$. The states with high energies $\omega>0$ have only slightly more spectral weight than the states with low energies $\omega<0$. Because the integrand on the r.h.s. of Eq.~(\ref{EQUthermo}) is weighted by $\omega$, the resulting thermoelectric response coefficient $X$ is small. Physically, this means that a small temperature difference between the leads induces the flow of high energy particles through the quantum dot, which, in principle, can result in a voltage drop across the junction. In the case $\gamma=0$ however, the current is compensated by a nearly equal counterflow of low energy carriers, so that the overall thermoelectric effect is small. If $\Delta$ is lowered to $1.8$, the low-energy states have the larger spectral weight and the thermoelectric response coefficient changes sign. For $\Delta=\varepsilon_p=2$, the spectral function is symmetric around $\omega=0$ so that the net charge current induced by the temperature difference vanishes and we have $X=0$.

For $\gamma=1$, the strong renormalization of the DL coupling reduces the width of the bands in the spectral function in Fig.~\ref{fig7}(c). As a result, near the Fermi edge the relative weight of the high-energy states increases, so that the dot acts as a more effective energy filter. The unfavorable counterflow of low-energy charge carriers is suppressed and the thermoelectric response $X$ grows considerably (see Fig.~(\ref{fig7}(b)). As can also be seen in Fig.~\ref{fig7}(b), the linear response conductance $L$ in Eq.~(\ref{EQUconductance}) decreases when $\gamma$ is set from zero to one, since it depends only on the (shrinking) spectral weight around the Fermi edge. This, too, boosts the thermopower $S$.

At $\Delta=2.2$ the variational calculation yields $\gamma_\mathrm{min}=0.5$, so that the width of the zero-phonon band lies between the other results. Consequently, this is also true for the maximum value of $X$. Note however, that our variational calculation maximizes the linear response conductance $L$ with respect to both limiting cases, so that the maximum thermopower is only slightly larger than for $\gamma=0$.
We conclude, that the local EP interaction can, in principle, enhance the maximum thermopower of the quantum dot device. Yet, for intermediate DL coupling strengths the small polaron picture with $\gamma=1$ greatly overestimates the effect. 
\section{Concluding remarks}
\label{sec:outlook}
To summarize, adopting a generalized variational Lang-Firsov transformation we calculate the interacting spectral function of a molecular quantum dot for small-to-large DL coupling and weak-to-strong EP interaction. We investigate the impact of the formation of a polaronic dot state on the steady-state current-voltage characteristics, as well as on the linear response thermopower of the system. 

In the case of strong EP interaction, the voltage-dependent polaronic renormalization of the DL coupling causes negative differential conductance. For comparable electronic and phononic time scales, this effect is diminished by transport through overlapping phonon side bands.

We find that in the off-resonant or ungated configuration, the renormalized dot level follows the lead chemical potentials. This process generates new peaks in the differential conductance signal.

In the equilibrium situation, the EP coupling enhances the thermopower of the quantum dot device, albeit by a smaller factor than predicted in the small polaron limit. 

The present work may be extended in several directions. Most notably, in the nonequilibrium regime, one should investigate the impact of the observed NDC on the thermoelectric properties of the molecular junction. The dynamics and heating of the vibrational subsystem could be included by means of nonequilibrium phonon Green functions~\cite{GNR07}. Moreover, in the light of recent advances in nanotechnology and experimental studies, new geometries have come into focus, like multi-terminal junctions or a molecule placed on an Aharonov-Bohm ring~\cite{EA11}.

%
%
\begin{acknowledgments}

This work was supported by Deutsche Forschungsgemeinschaft through SFB 652 B5. TK and HF acknowledge the hospitality at the Institute of Physics ASCR.

\end{acknowledgments}
%
%
%
%
%
\end{document}